\pdfoutput=1
\documentclass[utf8]{FrontiersinHarvard} % for articles in journals using the Harvard Referencing Style (Author-Date), for Frontiers Reference Styles by Journal: https://zendesk.frontiersin.org/hc/en-us/articles/360017860337-Frontiers-Reference-Styles-by-Journal
\usepackage{url,hyperref,microtype,subcaption}
\usepackage[onehalfspacing]{setspace}
\usepackage[normalem]{ulem}

\ifdefined\DeclareUnicodeCharacter\else
  \newcommand{\DeclareUnicodeCharacter}[2]{}
\fi
\DeclareUnicodeCharacter{2212}{-}
\DeclareUnicodeCharacter{2010}{-}
\DeclareUnicodeCharacter{2013}{--}
\DeclareUnicodeCharacter{2014}{---}
\DeclareUnicodeCharacter{00A0}{ }
\DeclareUnicodeCharacter{22C6}{*}
\DeclareUnicodeCharacter{2500}{-}
\DeclareUnicodeCharacter{2605}{*}
\def\keyFont{\fontsize{8}{11}\helveticabold }
\def\firstAuthorLast{E. Fiorellino and A. Somigliana}

\def\Authors{E.~Fiorellino\,$^{1,2*}$ and A.~Somigliana\,$^{3}$}

\def\Address{$^{1}$Alma Mater Studiorum – Università di Bologna, Dipartimento di Fisica e Astronomia “Augusto Righi”, Via Gobetti 93/2, I-40129,
Bologna, Italy \\
$^{2}$INAF – Osservatorio Astronomico di Trieste, via Tiepolo 11, I-34143 Trieste \\
$^{3}$Max-Planck-Institut für Astronomie, Königstuhl 17, 69117 Heidelberg, Germany}
\def\corrAuthor{Eleonora Fiorellino}

\def\corrEmail{eleonora.fiorellino@unibo.it}

\newcommand{\lsun}{\mbox{L}_\odot}
\newcommand{\rsun}{\mbox{R}_\odot}
\newcommand{\msun}{\mbox{M}_\odot}
\newcommand{\pab}{\mbox{Pa}\beta}
\newcommand{\brg}{\mbox{Br}\gamma}

\newcommand{\micron}{ \textmu m}

\newcommand{\lacc}{L_{\rm acc}}
\newcommand{\macc}{\dot{M}_{\rm acc}}
\newcommand{\lstar}{L_\star}
\newcommand{\mstar}{M_\star}
\newcommand{\rstar}{R_\star}
\newcommand{\teff}{T_{\rm eff}}
\newcommand{\lbol}{L_{\rm bol}}
\newcommand{\tbol}{T_{\rm bol}}
\newcommand{\lline}{L_{\rm line}}

\begin{document}
\onecolumn
\firstpage{1}

\title[Protostellar Accretion]{The Accretion Process on Protostars} 

\author[\firstAuthorLast ]{\Authors} %This field will be automatically populated
\address{} %This field will be automatically populated
\correspondance{} %This field will be automatically populated

\extraAuth{}% If there are more than 1 corresponding author, comment this line and uncomment the next one.
%\extraAuth{corresponding Author2 \\ Laboratory X2, Institute X2, Department X2, Organization X2, Street X2, City X2 , State XX2 (only USA, Canada and Australia), Zip Code2, X2 Country X2, email2@uni2.edu}

\maketitle

\begin{abstract}

%%% Leave the Abstract empty if your article does not require one, please see the Summary Table for full details.
%\section{}
%For full guidelines regarding your manuscript please refer to \href{http://www.frontiersin.org/about/AuthorGuidelines}{Author Guidelines}.

The process of mass accretion onto Young Stellar Objects (YSOs) plays a fundamental role in determining the final stellar mass and setting the initial conditions for planet formation. Despite its critical role, our understanding of accretion remains fragmented, particularly for what concerns the earliest, protostellar phases (Class 0/I). While the community has consolidated a comprehensive knowledge of the accretion process of the later-stage Classical T\,Tauri Stars (CTTSs), a similar level of understanding is critically lacking for the protostellar phase, where the bulk of the mass is assembled. This work aims to review recent major results, both from the observational and numerical point of view, bridging the gap between the two approaches and providing an updated, complete assessment of accretion in protostellar sources. We present different techniques to measure accretion on protostars, analyze how methodological differences affect parameter estimation, discuss the caveats in comparing with numerical models, and suggest the next steps to take towards an ever more exhaustive picture of the protostellar phase.

\tiny
 \keyFont{ \section{Keywords:} accretion, accretion disks – protoplanetary disks – stars: low-mass – stars: protostars} %All article types: you may provide up to 8 keywords; at least 5 are mandatory.
\end{abstract}

\section{Introduction}\label{sec:introduction}

Low-mass ($\leq 2\,\msun$) star formation is fundamentally regulated by the accretion process, which governs the assembly of the stellar mass. The term ``accretion" can refer to both \emph{stellar accretion}, the transfer of material from the inner disk onto the protostellar surface, or to \emph{disk accretion}, when the material is moving across or onto the disk at larger radii. In this review, we will focus on the first one, referring to it as accretion for brevity. 

The accreting material, stored in a surrounding envelope and/or an accretion disk orbiting the protostar at different evolutionary stages, approaches the protostellar surface as it loses angular momentum and migrates inwards; the actual mechanism leading to the mass incorporation depends on the strength and morphology of the magnetic field. At the protoplanetary disk stage, when the forming star is only surrounded by a disk of gas and dust, observations and theoretical models consistently support the magnetospheric accretion (MA) paradigm \citep{Hartmann2016, Manara2023}. Alternative magnetospheric configurations, such as the X-wind scenario, have also been proposed within this framework \citep[e.g.,][]{Shang2007}.
In this context, material from the infalling envelope is first processed through the circumstellar disk, where its inward transport is regulated by either, or a combination of, a macroscopic viscosity \citep{LyndenBellPringle1974, Pringle1981, shakura1973black-bc0} and magneto-hydrodynamical (MHD) winds \citep{Blandford1982, Lesur2020reviewMHD}. At the inner disk edge, the flow is typically truncated by the stellar magnetic field and channeled along magnetic field lines onto the stellar surface. 

Whether magnetospheric accretion also operates during the earlier protostellar phases, however, remains an open question. MA relies on specific physical conditions, most notably the presence of strong, large-scale stellar magnetic fields of the kilogauss level; when the magnetic field is weak, topologically complex, or unable to effectively truncate the disk, a different accretion regime may operate. The alternative scenario, boundary-layer (BL) accretion, produces high accretion rates where inflow pressure exceeds magnetic field pressure.  
In this picture, the inner disk is not truncated by the magnetospheric field lines but instead extends down to the stellar surface, allowing disk material to accrete directly onto the star; this mass transfer happens within a narrow boundary layer at the star–disk interface, rather than in magnetically funneled accretion shocks \citep{LyndenBell1974, Pringle1981, Popham1993, Popham1997}. Such conditions may be common during the earliest protostellar phases, when the magnetic fields are not developed or not strong enough yet and high accretion rates are needed to build the forming star \citep{Hartmann2016}. Nevertheless, current observational constraints do not rule out the presence of dynamically important magnetic fields in protostars - in particular for the less embedded ones; magnetospheric accretion remains therefore a physically viable and widely adopted framework even in the protostellar phases. Discriminating between these regimes is still an open theoretical and observational challenge.

Regardless of the exact physical mechanism, accretion is intrinsically linked to mass ejection phenomena, such as jets and winds, which play a crucial role in extracting angular momentum from the disk–star system and thereby enable continued mass accretion \citep[e.g.,][]{Blandford1982, Shu1994, Pudritz2007, Frank2014}. In particular, the X-wind model explicitly couples magnetospheric accretion and jet launching at the inner disk truncation radius \citep{Shu1994}; through this coupling, the accretion process regulates not only the transfer of mass onto the forming star, but also the angular momentum budget and the thermal and dynamical evolution of the inner disk.
The different accretion regimes, as well as other more distributed disk-mediated inflows (commonly referred to as {\it streamers}), are therefore expected to have distinct impacts on the structure, stability, and susceptibility to fragmentation of the inner disk; their influence is concentrated in the region at and within the co-rotation radius (typically a few stellar radii, corresponding to $\sim 0.02 – 0.1$ au in low-mass systems). %\sout{Moreover, the accretion history established during the embedded phases sets the initial conditions for planet formation and early planet–disk interactions. The large mass reservoir available in young disks, combined with episodic and potentially violent accretion events, may strongly influence the formation pathways of giant planets and the early accretion of planetary envelopes.} 
Moreover, the stellar and disk accretion history established during the embedded phases, characterized by gravitational instability and episodic bursts \citep[e.g.,][]{vorobyov2005origin-9cd, vorobyov2006burst-b61, DunhamVorobyov2012}, determines the mass distribution, thermal structure, and stability of young disks \citep[e.g.,][]{Kratter2010, Tsukamoto2017}; as a consequence, it sets the initial conditions for subsequent planet formation and early disk evolution. 
Observations indicate that Class 0/I disks can host substantial mass reservoirs \citep[e.g.,][]{Tobin2016, tyc18, SheehanEisner2017}, while theoretical studies show that massive, gravitationally unstable disks may undergo fragmentation, 
potentially leading to the formation of giant companions \citep[e.g.,][]{Kratter2010, ForganRice2011, Vorobyov2013}. 
In addition, episodic and potentially violent accretion events can significantly alter the thermal structure of the disk \citep[e.g.,][]{bae2014accretion-224, Meyer2017}, thereby influencing both fragmentation conditions, the early growth of massive planetary envelopes, and the physical and chemical conditions of the disk material \citep{Morbidelli2024}.

In this context, identifying which accretion mechanisms operate during the protostellar phase, and quantifying their intensity, geometry, and efficiency, is of paramount importance to understand the stellar mass budget and to link star formation, disk evolution, and the emergence of planetary systems.
Over the past decade, several comprehensive reviews have addressed the physics of accretion in young stellar objects. However, most of them have primarily focused on the more evolved young stars, where magnetospheric accretion is observationally well established \citep[e.g.,][]{Hartmann2016, Manara2023}; in these works, the protostellar stages are typically only briefly mentioned, mainly in the context of the early mass assembly or episodic accretion. Conversely, the most recent dedicated review of the protostellar phase \citep{Dunham2014} provides a broad overview of embedded evolution in the aftermath of the major \textit{Spitzer} and 2MASS surveys, but devotes limited space to the detailed physics and diagnostics of accretion itself.
Since then, the field has undergone substantial progress; new observational facilities and techniques — particularly high-resolution infrared spectroscopy and interferometry with instruments such as at the Very Large Telescope (VLT) and, more recently, the James Webb Space Telescope (JWST) — have enabled direct probes of embedded accretion processes that were previously inaccessible. At the same time, advances in numerical simulations have refined our understanding of disk instability, episodic accretion, magnetic regulation, and disk–envelope interaction during the embedded phases. Given these developments, and in view of the transformative capabilities expected from forthcoming facilities such as the Extremely Large Telescope (ELT), a comprehensive reassessment of low-mass protostellar accretion is timely. This review aims to provide such an update, synthesizing the observational and theoretical progress achieved over the last decade and outlining the key open questions that will shape the next generation of studies.

With this goal in mind, we have structured this review as follows. We describe the classification of Young Stellar Objects in Section\,\ref{sect:classification}; we summarize the current constraints on protostellar accretion from the observational and theoretical perspective in Sections\,\ref{sect:observations}\,and\,\ref{sec:models_and_sims}, respectively; we discuss the emerging picture of protostellar accretion in Section\,\ref{sect:main-res-open-questions}, and finally outline the open questions and next challenges in Section\,\ref{sect:conclusions}.

\begin{figure}
    \centering
\includegraphics[width=1.0\linewidth]{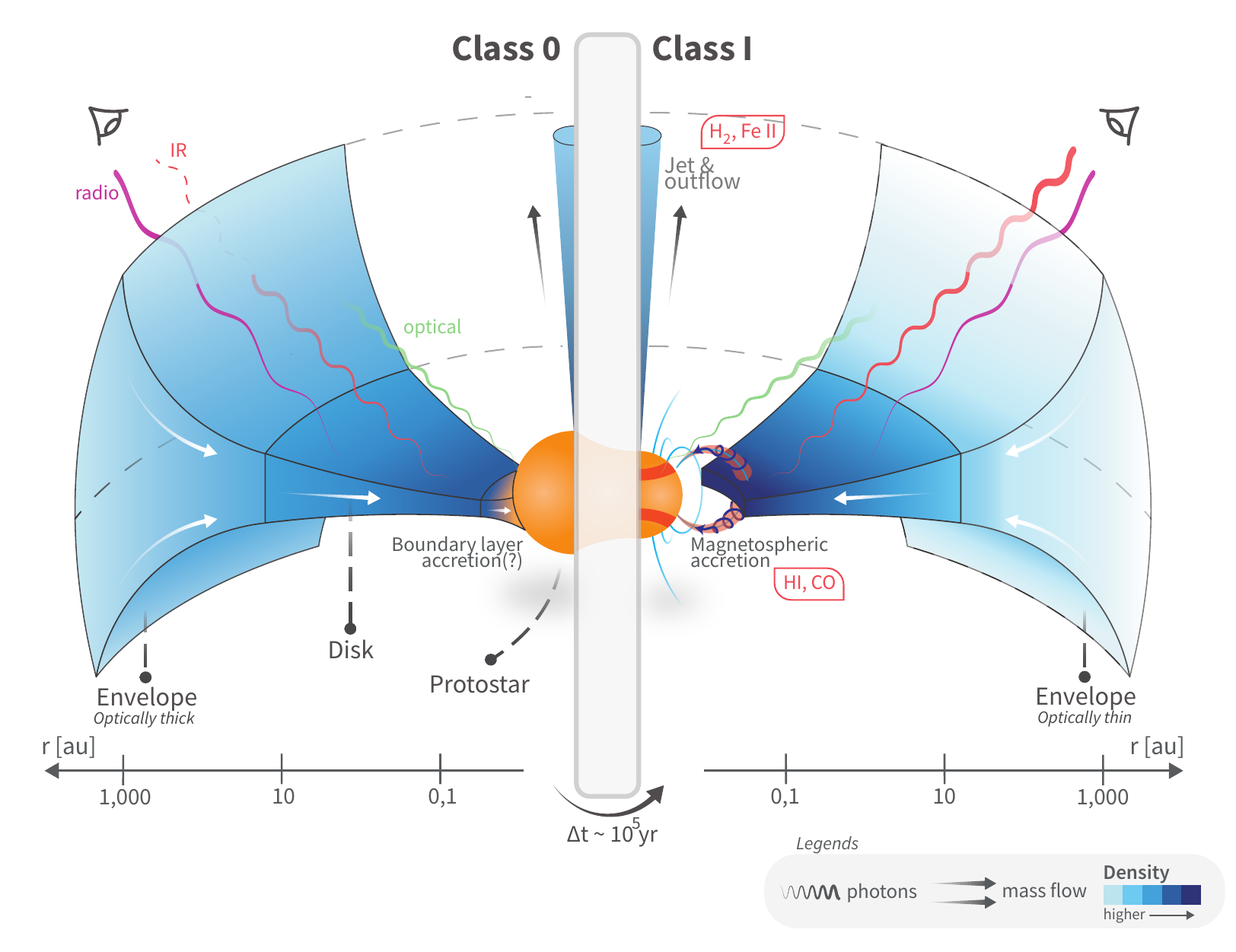}
    \caption{Schematic representation of the accretion process in Class\,0 and Class\,I protostars. Class\,0 objects (left) are characterized by a massive, optically thick envelope dominating the system mass and feeding the disk and protostar, with accretion possibly occurring through non-magnetospheric inner-disk processes. In Class\,I sources (right), the envelope becomes progressively thinner and less massive, the disk plays a more central role, the magnetic field get stronger, and magnetospheric accretion may emerge, producing accretion diagnostics such as H{\footnotesize I} and CO emission. The figure summarizes the evolution of density structure, mass-flow geometry, and observational tracers across the protostellar phase. In both stages, the accretion process is expected to be accompanied by powerful jets and outflows, traced by shock-excited species such as H$_2$ and [Fe{\footnotesize II}], which play a key role in extracting angular momentum from the system. Figure credits: Dr. Adrien Houge.}
    \label{fig:class0I}
\end{figure}

\section{Classification of Accreting Young Stellar Objects}
\label{sect:classification}

Star formation begins with the gravitational collapse of prestellar cores, overdense structures within molecular clouds often associated with filamentary environments \citep[e.g.,][]{Arzoumanian2011, Andre2014, Konyves2015}. When the core mass exceeds the Bonnor--Ebert critical mass, gravitational collapse ensues \citep[e.g.,][]{Ebert1955, Bonnor1956, KetoCaselli2008}; during the collapse, the increasing density and opacity of the gas lead to the formation of the first hydrostatic core, a transient, pressure-supported object in which the collapse is temporarily halted \citep{Larson1969, MasunagaInutsuka2000}. The onset of the second collapse then gives rise to the formation of a protostar surrounded by a circumstellar disk, marking the beginning of the protostellar phase, during which the forming star remains embedded within its parental envelope and accretes a substantial fraction of its final mass \citep{Stahler1986}. By the time the envelope dissipates, most of the stellar mass has been assembled, the protostellar phase ends, and the young star enters the pre--main-sequence (PMS) phase. In this Section, we summarize the main evolutionary stages of Young Stellar Objects (YSOs), describe their corresponding observational classes, and discuss the limitations of this classification.

Historically, YSOs have been observationally divided in Classes based on ({\it i}) their infrared (IR) spectral index, computed between 2 and 24\micron\ as $\alpha_{\rm IR} = d \log(\lambda F_\lambda)/d \log(\lambda)$ \citep{Lada1987}, ({\it ii}) their bolometric temperature $\tbol$ \citep{MyersLadd1993}, and ({\it iii}) the ratio of submillimeter to bolometric luminosity $L_{\rm submm}/\lbol$, where $L_{\rm submm}$ is defined as the integrated luminosity at wavelengths $\lambda \geq 350$\micron\ \citep{and93}. The spectral index $\alpha$ has also been computed in the $4.5 - 24$\micron\ wavelenght range, since the 4.5\micron\ {\it Spitzer} channel is less sensitive to extinction than 2\micron\ \citep{Kryukova2012, Dunham2014, Furlan2016}. The commonly adopted mapping between observational Classes and evolutionary stages of accreting young stars can be summarized as follows:

\begin{itemize}
    \item \textbf{Class\,0} (Early protostellar phase): These sources represent the earliest stage of star formation. They are deeply embedded within their parental envelopes (optically thick regime), and are believed to have assembled only a small fraction of their final stellar mass. They are characterized by $\alpha \geq 0.3$ and $\tbol < 70$\,K. Statistical studies suggest Class\,0s to have lifetimes of $\sim 10^4$--$10^5$\,yr \citep[e.g.,][]{Andre2000, Evans2009, Dunham2014}. The mass accretion rate is expected to be very high during this stage, and episodic accretion bursts are thought to be intense and frequent. %We also note that in the HOPS/eHOPS SED classification, $\tbol$ were used to distinguish Class\,0 from Class\,I and then $\alpha$ between 4.5 to 24\,$\mu$m to distinguish FS protostars from Class\,I and Class\,II \citep{Furlan2016, Pokhrel2023}. 

    \item \textbf{Class\,I} (Later protostellar phase): Class\,I sources are protostars still embedded in substantial envelopes, which are however more optically thin than those around Class\,0 objects. They have $\alpha \geq 0.3$ and $70 < \tbol < 650$\,K; whether the bulk of the stellar mass is already assembled at this stage remains uncertain. The Class\,I phase is estimated to last a few $\times 10^5$\,yr \citep[e.g.,][]{Evans2009, Dunham2014}; like for Class 0s, they are likely to have very high mass accretion rates, as well as frequent, intense episodic accretion bursts.

    \item \textbf{Flat-spectrum} ({\bf FS}) (Transition between phases): FS sources are generally interpreted as a transitional population between the protostellar and PMS phases. They exhibit $-0.3 < \alpha \leq 0.3$ and $650 < \tbol \leq 2800$\,K and are likely to have assembled most of their final stellar mass. Their typical lifetime is estimated to be of order $\sim 10^5$\,yr, although with large uncertainties \citep[e.g.,][]{Greene1994, Evans2009}. It has been suggested that FS can either have low density envelopes or low inclination with denser envelopes \citep{Calvet1994, Habel2021, Federman2023}; the mass accretion rate is relatively low and episodic bursts are rare.

    \item \textbf{Class\,II} (PMS phase): Class\,II objects, also known as Classical T\,Tauri Stars (CTTSs), are PMS stars surrounded by circumstellar disks but largely devoid of envelopes. Their Spectral Energy Distributions (SEDs) are characterized by $\alpha \leq -0.3$ and $\tbol > 2800$\,K, and their stellar masses are essentially set (although they are still actively accreting material from the disk around them). The Class\,II phase typically lasts $\sim 2$--$3$\,Myr \citep[e.g.,][]{Haisch2001, Evans2009}.

    \item \textbf{Class\,III} (PMS phase): Class\,III sources represent the latest stage of the star formation process. They have dispersed most of their circumstellar disks and show little to no infrared excess, corresponding to weakly accreting or non-accreting young stars. This stage extends over several Myr, with typical timescales of $\gtrsim 5$--$10$\,Myr \citep[e.g.,][]{Haisch2001, Fedele2010}.
\end{itemize}

Figure\,\ref{fig:class0I} provides a summary of the main observational and physical properties of Class\,0 and Class\,I sources. Throughout this review, we use the term \textit{protostars} to refer to Class\,0, Class\,I, and FS objects, specifying the individual Class when necessary. We frequently compare results obtained for protostars with those derived for the more extensively studied CTTS population, partly because CTTSs are less embedded and suffer lower extinction than protostars, making them more observationally accessible.

\begin{figure}
   \centering
\includegraphics[width=0.5\linewidth]{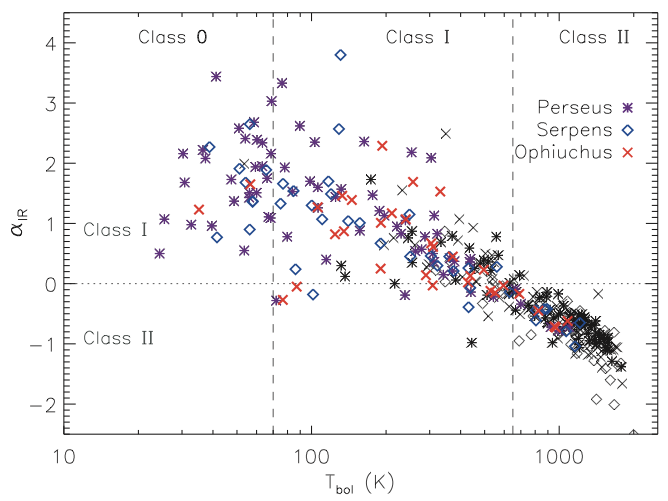}
  \caption{Spectral index vs. bolometric temperature for young stars analysed with {\it Spitzer} in Perseus, Serpens, and Ophiuchus. The bold symbols indicate sources that are associated with 1.1 mm emission, and the thin symbols denote those with upper limits at 1.1 mm. Class divisions for both $\tbol$ and $\alpha_{\rm IR}$ are shown. The two methods agree fairly well for Class\,II and “warmer” Class\,I sources, but very cold ($\tbol \lesssim 100$\,K) sources have a large range of $\alpha_{\rm IR}$ values. Adapted from \cite{eno09}. © AAS. Reproduced with permission.}
   \label{fig:Enoch+2009}
\end{figure}

Ideally, classifications based on these different diagnostics would yield consistent results. In practice, however, discrepancies are common. According to \citet{eno09}, classifications based on $\tbol$ are more reliable for distinguishing protostars (Class\,0 and I) from PMS stars (Class\,II), as sources with $\tbol \leq 600$\,K are classified as Class\,I based on their bolometric temperature, yet often exhibit infrared spectral indices consistent with Class\,II. However, \citet{eno09} also show that sources with discrepant classifications predominantly occupy the range $-0.3 \leq \alpha_{\rm IR} \leq 0.3$, characteristic of flat-spectrum (FS) objects, which were introduced as a transitional class between Class\,I and II \citep{Greene1994} (see Figure \ref{fig:Enoch+2009}). 
Note that, instead, discrepancies in the classification become negligible when $\alpha$ is computed using 4.5\micron\ \citep{Furlan2016}.

Since the classification described above is based on observational diagnostics, caution is required when applying its vocabulary to numerical simulations or when directly associating observational Classes with physical evolutionary stages. We discuss the caveats and uncertainties in the following Section (\ref{sect:class-lifetime}). 

\subsection{Classification Uncertainties and Evolutionary Timescales}
\label{sect:class-lifetime}

The straightforward interpretation of observational Classes in terms of evolutionary stages, while tempting, requires caution: the derivation of evolutionary timescales from the SED diagnostic features is tightly linked to the definition and identification of Classes, which may result in a major cause of systematic error in lifetime estimates. Timescales associated with the different protostellar and PMS phases are a critical ingredient for understanding when and how stellar and planetary masses are assembled: in particular, estimates of the lifetimes of sources in the various Classes underpin our interpretation of mass accretion histories and set the temporal framework for disk evolution and planet formation. 

To date, the most widely adopted lifetimes are derived from population statistics based on IR photometric classifications, primarily from \textit{Spitzer} surveys \citep[e.g.,][]{Evans2009}. While this approach provides a homogeneous and statistically robust framework, it implicitly assumes that photometric Class assignments accurately trace evolutionary stage. However, growing observational evidence suggests that photometric diagnostics alone may be insufficient to unambiguously determine the evolutionary status of YSOs. Near-IR (NIR) spectroscopic follow-ups of \textit{Spitzer}/2MASS-selected samples have revealed that a non-negligible fraction of sources classified as protostars exhibit spectroscopic properties commonly associated with more evolved objects, including detectable photospheric absorption features and accretion signatures typical of CTTSs \citep{Fiorellino2021, LeGouellec2024}. A further illustrative case is that of HOPS\,315, identified as a Class\,I source based on its bolometric temperature ($T_{\mathrm{bol}} \sim 180$ K), but showing a molecular jet typically associated with younger objects, suggesting that it may instead be a Class\,0 protostar \citep{Dutta2022}. 
These findings provide a concrete example of the classification ambiguities and highlight the limitations of relying exclusively on SED-based criteria to obtain reliable evolutionary stage estimates. Another potential source of misclassification is the viewing angle. Geometrical effects can significantly distort the observed spectral energy distribution: for example, edge-on Class\,II sources may exhibit high extinction due to the disk midplane and be misidentified as Class\,I objects, while a Class\,I source viewed through an outflow cavity may appear as a Class\,II \citep{masunaga2000radiation-02f, robitaille2006interpreting-8ed}; there have also been hints of some Class\,II PMSs showing evidence for envelopes \citep{Furlan2016}. Accretion variability can further cause YSOs to temporarily cross class definition boundaries \citep{dunham2010evolutionary-7c4}.

In this context, \citet{Fiorellino2021} proposed that classification schemes based on spectroscopic diagnostics, rather than on the infrared spectral index alone, may offer a more physically meaningful description of YSOs, particularly for Class\,I and FS sources. This approach naturally links the classification to the dominant physical processes at play, including the emergence of stellar photospheres, accretion activity, and circumstellar obscuration. A complementary perspective is provided by \citet{Federman2023}, who demonstrated that the ratio of compact (disk) to extended (envelope) 870\micron\ emission -- measured with the Atacama Compact Array (ACA) and Atacama Large Millimeter Array (ALMA) out to 2500\,au -- can serve as an evolutionary indicator that is less sensitive to inclination and extinction effects than traditional SED-based classifications. Importantly, this metric does not redefine the Class\,0/I/FS categories, but rather provides an independent structural tracer that statistically correlates with them.
In their sample, Class\,0 protostars are predominantly envelope-dominated (with a disk-to-envelope flux ratio $\mathbf{R < 0.5}$), FS sources are largely disk-dominated ($\mathbf{R > 0.5}$), and Class\,I objects span both regimes ($\mathbf{R \in (0, 1)}$), suggesting that structural measures of the disk--envelope configuration can robustly trace the evolutionary progression.

Adopting such physically motivated classification metrics would likely lead to a reassessment of the relative populations of protostellar classes derived from \textit{Spitzer} surveys, and consequently of their inferred lifetimes. For instance, in the NGC\,1333 sample analyzed by \citet{Fiorellino2021}, 7 out of 17 objects ($\sim$40\%) that were identified as Class\,I based on their SEDs were found to be more consistent with Class\,II sources when spectroscopic diagnostics were considered. This shows that class lifetimes could be revised at a comparable level; in this framework, the traditional boundaries between Classes become less rigid, and evolutionary stages are instead defined by the physical structure of the system rather than by photometric slopes alone. Such revisions would directly impact estimates of the duration of the main mass-assembly phases, both for stars and for forming planetary systems.

When observational classes are implicitly interpreted as proxies for age or evolutionary stage, the very definition of ``protostar'' can be ambiguous. While the classification is formally based on IR spectral indices and bolometric temperatures, quantities that primarily trace the presence and properties of circumstellar envelopes, the terminology is often used interchangeably with chronological age. In this simplified interpretation, objects younger than $\sim 10^4$ yr are usually referred to as Class\,0; Class\,I/FS corresponds to ages between $10^4$ and $10^6$ yr; finally, Class\,II refers to ages older than $10^6$ yr. 
In order to disentangle observational appearance from physical structure, the term Stage has been introduced to denote a classification based on intrinsic properties -- such as the relative mass of the envelope and disk or the envelope accretion rate -- rather than on spectral indices \citep[e.g.,][]{Andre1993, Robitaille2006}. In this framework, Stage 0/I sources are defined by the presence of a substantial infalling envelope (e.g., $M_{\rm env} \gtrsim M_\star$ or $\dot{M}_{\rm env}/M_\star$ above a given threshold), independently of their observed SED shape.
Although the Stage classification is physically motivated, mapping observational Classes onto physical Stages remains non-trivial.
The age-based interpretation is intuitively appealing, however conflating observational class with evolutionary time can be problematic, particularly when comparing objects of very different stellar masses (for example, at a fixed age, a 1\,$\msun$ star is significantly more evolved than a 0.1\,$\msun$ star, since PMS contraction timescales are strongly mass-dependent). 

More issues arise when comparing observations with numerical simulations; in the latter, the definition of ``age'' is often model-dependent and may refer to different reference points, such as the onset of core collapse, the formation of a protostar, or the attainment of a specific physical condition. Moreover, simulations that begin with pre-existing stellar objects may not explicitly account for the earlier collapse and accretion phases. 
For these reasons, caution is required when directly mapping observational classes onto simulated evolutionary timelines. A more robust alternative to pinpoint the Class of a simulated YSO is based on the envelope mass fraction $M_{\rm{env, f}}$, defined as the ratio of the envelope mass over total (envelope + sink) mass; in this framework, Class\,0 corresponds to $M_{\rm{env, f}} > 0.5$, while Class\,I is $0.1 < M_{\rm{env, f}} < 0.5$. Classification schemes based on spectral diagnostics or on physically motivated flux ratios may also provide a more direct link to the quantities tracked in numerical models, thereby facilitating a more meaningful comparison between observations and simulations. 
Observationally, estimates of the central protostellar mass based on kinematic measurements of disks and envelopes are becoming increasingly available for well-studied sources, providing an important empirical anchor for such physically motivated classifications (see Section\,\ref{sect:stellar_par}). We emphasize this caveat not to discourage such comparisons, but rather to encourage careful and consistent interpretations.

\section{Observations}\label{sect:observations}

%\subsection{Measurements of protostellar accretion parameters} \label{section:Lacc}

Because protostars are deeply embedded within their natal envelopes, optical observations are severely limited. IR and radio wavelengths, where the radiation from the forming star is reprocessed and re-emitted by dust in the disk and envelope, are better suited for observing these objects; however, not all IR (especially NIR) emission arises from purely reprocessed radiation. At wavelengths around $\sim$2\micron\, part of the observed flux may consist of direct or scattered photospheric emission, as well as hydrogen recombination lines tracing accretion, as a consequence of the local reprocessing of the UV/optical photons or their observation in scattered light through outflow cavities \citep[e.g.,][]{Habel2021}.
Observations at these wavelengths provide access to a range of diagnostics that can be used to infer the fundamental stellar properties and the characteristics of the accretion process. 

In this Section, we present the main stellar and accretion parameters considered in this review; after an overview of the relevant quantities, we describe each of them in depth, outlining the observational diagnostics and techniques commonly employed to measure them.

The accretion luminosity ($\lacc$) quantifies the energy released as material accretes from the circumstellar disk onto the forming star. 
For CTTSs, $\lacc$ has been constrained for large samples by modeling the ultraviolet (UV) continuum excess emission generated by accretion shocks above the stellar photosphere \citep[][]{calvet1998, Schneider2020, pittman2022, Manara2023}, as well as through empirical correlations between $\lacc$ and the U-band excess luminosity \citep[][]{val93, Gullbring1998, Herczeg2008, Sicilia-Aguilar2010}. Unfortunately, applying these methods to embedded protostars is challenging. 
In these objects, the U-band emission is severely attenuated by the high visual extinction ($A_V$) associated with the dense envelope; furthermore, strong continuum veiling ($r$) requires the detection of the stellar photosphere, which is often impossible because of the faintness of the photospheric emission compared to the accretion luminosity. 
As a result, direct modeling of the UV excess becomes impractical, and constraining $\lacc$ in the earliest protostellar phases remains highly challenging.

Alternatively, the accretion luminosity can be derived from emission lines luminosity ($\lline$), such as H\,{\footnotesize I}, He\,{\footnotesize I}, Fe\,{\footnotesize I}, Ca\,{\footnotesize I}, Na\,{\footnotesize I}, and O\,{\footnotesize I}, using empirical relations between $\lacc$ and $L_{\rm line}$ \citep{muz98, Herczeg2008, alc14, alc17, Fiorellino2025}. 
It is important to stress, however, that a tight $\lacc$–$L_{\rm line}$ correlation does not necessarily imply a direct physical or causal link between the line-emitting region and the accretion shock itself; as discussed by \citet{Mendigutia2015}, part of the observed correlations may arise from underlying dependencies on stellar luminosity or other global parameters, rather than from a one-to-one physical connection between line formation and accretion energy release. Moreover, interferometric studies have shown that hydrogen recombination lines such as Br$\gamma$ and H$\alpha$ do not always originate exclusively in magnetospheric funnel flows, but can also trace disk winds or more extended inner-disk regions, while still preserving the empirical $\lacc$–$L_{\rm line}$ calibration. Similarly, there has been evidence of empirical relations between $\lacc$ and the luminosity of forbidden lines such as [O\,{\footnotesize I}]$\lambda$6300, whose origin is clearly associated with outflows rather than direct accretion. These examples highlight that an empirical calibration should not be automatically interpreted as evidence for a specific accretion geometry or emission mechanism.

Building on this approach, and making the critical assumption that the empirical $\lacc$--$L_{\rm line}$ relations originally calibrated for CTTSs are applicable to embedded protostars under the MA scenario, the study of accretion in these young objects has progressed significantly in recent years. Notably, \citet{whi04} were able to study a sample of 15 very bright Class\,I protostars at optical wavelengths, concluding that most of the sources in their sample are not in the main accretion phase and are older than previously assumed.
This was possible because the targets were among the optically brightest Class\,I sources, likely characterized by relatively low extinction and favorable viewing geometries. 
Such optical analyses, however, remain the exception. 
In most cases, dust in the protostellar disk and envelope absorbs the stellar and accretion luminosity at optical wavelengths and re-emits it at longer wavelengths; as a consequence, protostars are typically very bright in the IR and largely inaccessible at shorter wavelengths.
At NIR bands, in particular, emit a number of commonly used tracers which provide a powerful tool to investigate both stellar and accretion properties of protostars. The main ones are (i) strong H\,{\footnotesize I} lines tracing accretion, such as $\pab$ and $\brg$, formed in the hot ($T \sim 10^4$\,K), dense gas of accretion columns; (ii) CO band-heads in the $K$-band, which probe the warm inner disk region; (iii) wind and jet tracers, such as [Fe\,{\footnotesize II}] at 1.64\micron\ and H$_2$ lines at 2.12 and 2.24\micron, which are also indirectly linked to accretion; (iv) stellar photospheric features, which allow spectral-typing and stellar parameters constraining; and finally (v) the $J$-band magnitude, which can be used to estimate the stellar luminosity ($\lstar$).

Under the assumption that accretion onto protostars proceeds via magnetospheric accretion, as in CTTSs, stellar parameters such as the stellar radius ($\rstar$) and mass ($\mstar$) are required not only to gain a deeper understanding of the protostellar phase, but also to compute the mass accretion rate:

\begin{equation}\label{eq:01}
\macc \sim \left( 1 - \frac{\rstar}{R_{\rm in}} \right)^{-1} \frac{\lacc \rstar}{G \mstar},
\end{equation}

where $G$ is the gravitational constant and $R_{\rm in}$ is the inner radius, usually assumed to be $\sim 5 \,\rsun$ for CTTSs \citep{Gullbring1998}. 
Direct NIR interferometric measurements with VLTI/GRAVITY have spatially resolved the $\brg$ emitting regions in several young CTTS systems, finding characteristic sizes of only a few stellar radii. 
In DoAr 44, the $\brg$ region is constrained to an upper limit of $\sim$0.047\,au ($\sim 5\,\rstar$, \citealt{Bouvier2020}), and similarly compact sizes ($\sim4.8\,\rstar$) are reported for CI Tau in the GRAVITY YSO survey \citep{Gravity2023inner-disk}. 
These spatial scales are broadly consistent with expectations for magnetospheric truncation radii in low-mass accreting stars. Although these observations trace the spatial extent of the $\brg$ emission rather than the magnetosphere directly, the inferred sizes provide an observational estimate of the inner disk radius $R_{\rm in}$, commonly associated with the magnetospheric truncation radius in the magnetospheric accretion framework.
In addition, recent spectroscopic modeling suggests that the inner disk radius in CTTSs may lie as close as $\sim 2-3 \, \rstar$, smaller than the canonical $\sim 5\,\rstar$ often assumed in magnetospheric accretion prescriptions \citep{Pittman2025}. Together, these results indicate that magnetospheric truncation radii are likely confined to a few stellar radii, though with significant object-to-object variation. Equation\,\eqref{eq:01} implicitly assumes radiatively efficient (cold) accretion, whereby the gravitational potential energy released by the infalling material is entirely radiated away. If a fraction of the accretion energy is instead absorbed by the protostar (hot accretion), the inferred mass accretion rates would be correspondingly higher.
The applicability of this assumption to embedded protostars is discussed in Section\,\ref{sect:accretion-nature}.

Lastly, in protostellar systems the disk is not an isolated structure, but interacts with both the forming star (by providing the material that is accreted) and the surrounding envelope (from which it is continuously replenished). 
As a result, the accretion process dynamically links the disk mass to the growth of the central protostar; disk and envelope masses and sizes therefore carry key information on the accretion process and on the efficiency and timescales of mass assembly during the embedded phases.

In the following subsections, we describe the main parameters required to study the accretion process and the spectroscopic features linked to them. For a discussion on the applicability of the magnetospheric accretion assumption during the protostellar phase, see Section\,\ref{sect:main-res-open-questions}.

\subsection{Stellar parameters} \label{sect:stellar_par}

We refer to stellar parameters as the quantities describing the intrinsic properties of a protostar, such as the effective temperature ($\teff$), surface gravity ($\log g$), stellar luminosity ($\lstar$), radius ($\rstar$), and mass ($\mstar$). 
We include in this list also the visual extinction ($A_V$) measured along the line of sight of the forming star; this is because, despite not being an intrinsic property of the protostar, its accurate estimate is required to reliably derive these parameters from observations.

Accurate spectral typing, obtained by comparing the photospheric absorption lines of protostars with those of non-accreting Class\,III templates or with synthetic stellar atmosphere spectra, allows the determination of $\teff$, $A_V$, and $\log g$, provided that spectra of sufficiently high resolution and signal-to-noise are available.
From short bands, ideally optical or the $J$ band, $\lstar$ can be estimated using appropriate bolometric corrections; once placed on the Hertzsprung--Russell diagram, evolutionary tracks can then be used to derive $\rstar$ and $\mstar$.
While this is not the only method to determine stellar parameters, it is the most commonly adopted in the literature and, unless otherwise stated, it is the method used in the works reviewed here. Indeed, based on this approach, high-resolution NIR spectra ($R \simeq 18{,}000$) from one of the first surveys targeting stellar properties of Class\,I and FS sources allowed spectral typing for 41 out of 52 objects \citep{dop05}. That analysis provided estimates of $\teff$, $\log g$, and $\lstar$ broadly similar to those of CTTSs, but also showed that large uncertainties, primarily due to the difficulty in constraining the extinction, prevented firm conclusions. 
In addition, higher projected rotational velocities ($v \sin i$) and angular momenta than those observed in Class\,II sources were reported.
In a subsequent study, doubling the sample size to 110 Class\,I and FS young stellar objects, \citet{con10} found a much wide range of extinction values ($A_V = 0$--$60$\,mag) and spectral types spanning from A0 to M7, with two main peaks around K2 and M3. These results suggest systematically later spectral types compared to CTTSs.

The protostellar mass can also be inferred using interferometric observations. Reaching angular resolutions of $\sim$0.05", corresponding to $\sim$10\,AU in nearby star-forming regions, ALMA\footnote{Atacama Large Millimeter Array} routinely resolves the velocity fields of young star disks. 
The stellar mass of protostars can be estimated from ALMA observations by modeling the rotation of molecular gas under the assumption of Keplerian motion, whereby the gas dynamics is dominated by the gravitational potential of the central object. In this framework, the azimuthal velocity of the gas is given by
$ v_\phi(r) = \sqrt{G M_\star/d_r}$, where $d_r$ is the radial distance from the central source. 
This approach has proven to be robust for Class\,II systems, where the circumstellar disk is well defined and largely isolated, and the observed velocity field is typically consistent with purely Keplerian rotation \citep[e.g.,][]{Simon2000, Pinte2018}. In this case, stellar masses can be derived with uncertainties of the order of $\sim$5--10\%, increasing to $\lesssim$20\% when uncertainties in disk inclination are taken into account.

For more embedded sources, however, the validity of the Keplerian assumption becomes increasingly uncertain. In Class\,I protostars, the disk is often still embedded within a rotating and infalling envelope, whose contribution to both the gravitational potential and the observed velocity field can lead to deviations from purely Keplerian rotation \citep{Tobin2012, Harsono2014, Yen2015, Aso2015}. 
As a result, the measured velocities may reflect a combination of disk rotation, envelope infall, and projection effects, and the derived masses are frequently better described as dynamical masses that include contributions from the inner disk and envelope. 
Reliable $\mstar$ estimates in Class\,I systems therefore require high angular resolution to spatially isolate the disk-dominated region, as well as the use of optically thin molecular tracers. Even under favorable conditions, the determination of stellar masses carries significant uncertainties, with typical values of order $\sim$20--50\% \citep{Yen2017, Segura-Cox2020}.

In Class\,0 protostars, the situation is even worse as the circumstellar environment is generally dominated by a massive envelope, and clear Keplerian disks are either absent or confined to very small radii. 
Even when compact disks are present, identifying Keplerian rotation is challenging not only because of limited spatial resolution, but also because of high optical depths and line of sight contamination from the infalling envelope, which can obscure or distort the kinematic signature of the inner disk. In these systems, the gas kinematics are often governed by infall and angular momentum conservation rather than by Keplerian rotation, making direct stellar mass measurements highly uncertain or unfeasible \citep{Tobin2012, Ohashi2014, Yen2015, Maury2019}, with one of the higher sources of uncertainty being how the velocity of the disk near the protostar is measured \citep{Tobin2024}. Consequently, stellar masses inferred from ALMA observations of Class\,0 sources are often reported as upper or lower limits, or as total dynamical masses rather than as direct measurements of the protostellar mass alone.

Although dynamical stellar masses cannot generally be derived for embedded protostars with the same accuracy and reliability as for Class\,II systems, they provide independent physical constraints that can be used to reduce the large uncertainties affecting $\mstar$ inferred from evolutionary tracks, for instance by ruling out incompatible solutions or by providing meaningful upper limits on $M_\star$.

Alternatively, stellar parameters can be derived by fitting the observed SED with theoretical models \citep[e.g.,][]{rob06, Hatchell2007}. This approach relies on a strong theoretical framework and requires a well-sampled SED, ideally constructed from simultaneous multi-wavelength observations in order to minimize the effects of variability. In practice, however, SED fitting primarily constrains the total luminosity that heats the disk and envelope, and separating $L_{\rm acc}$ from $L_\star$ is highly model-dependent and often degenerate. In many Class\,0 applications, the heating luminosity is effectively assumed to be dominated by $L_{\rm acc}$; moreover, even with a perfectly sampled SED this method also requires to know what fraction of the luminosity is due to accretion and to adopt a mass-to-radius relationship, which in most cases impacts the accuracy of this methodology. This approach is used when high-quality observations at the appropriate wavelengths are not available, and is particularly common for Class\,0 sources, for which direct constraints on stellar parameters from spectroscopy or disk kinematics are often unavailable.

In summary, despite significant progress in the last decade, the accuracy with which stellar and accretion parameters can be constrained in embedded protostars remains limited. For Class\,I sources, typical uncertainties are of order $\sim$1 spectral subclass in spectral type, $\sim$1\,mag in $A_V$, and up to $\sim$1\,dex in both $M_\star$ and $R_\star$ when derived from evolutionary tracks. While $\lacc$ can often be constrained within $\sim$0.3\,dex under favorable assumptions (see Section \ref{sect:empirical-relations}), the propagated uncertainty on $\dot M_{\rm acc}$ may reach $\sim 1 - 2$\,dex, owing to the combined uncertainties on extinction, veiling, stellar parameters, and the assumed accretion geometry.
For Class\,0 protostars, where direct spectroscopic constraints on stellar parameters are generally unavailable and disk kinematics are often dominated by envelope infall, the uncertainties are comparable or larger. In many cases, stellar masses and radii cannot be measured directly and are instead inferred through model-dependent assumptions or reported as upper or lower limits. As a consequence, both stellar and accretion parameters estimates for Class\,0 sources remain subject to substantial systematic uncertainties.

\subsection{Veiling}
\label{sect:veiling}

Veiling refers to the phenomenon whereby photospheric absorption lines appear weaker than expected in the stellar spectrum. This effect arises because an additional continuum emission is superimposed on the photospheric spectrum, partially filling in the absorption lines and making them appear shallower.
Quantitatively, the veiling at a given wavelength $\lambda$ is defined as
\begin{equation}
r_\lambda = \frac{F_{\rm excess}}{F_{\rm photosphere}},
\end{equation}
where $F_{\rm excess}$ is the excess continuum flux and $F_{\rm photosphere}$ is the intrinsic photospheric flux at wavelength $\lambda$.

In accreting systems, the excess continuum responsible for veiling is produced by the release of accretion energy. Under the assumption of magnetospheric accretion, gas from the circumstellar disk is funneled along stellar magnetic field lines and impacts the stellar surface at near free-fall velocities. 
This process generates accretion shocks with characteristic temperatures of $\sim 8{,}000$\,K, emitting primarily in the UVB and optical bands, as well as even hotter components contributing to the X-ray emission. 
At longer wavelengths, particularly in the NIR, a significant fraction of the continuum excess may instead originate from the hot inner disk rim at the dust sublimation radius, which is heated by stellar irradiation and accretion luminosity \citep[e.g.,][]{Muzerolle2003}.
These accretion hot spots produce a largely featureless continuum; as a result, veiling arises because the accretion-generated emission is broad-band, contains few or no absorption features, peaks in the UV and optical, and adds flux on top of the stellar spectrum, thereby reducing the contrast of photospheric absorption lines.
Consequently, higher accretion rates lead to stronger excess continuum emission and increased veiling, whereas lower accretion rates result in weaker or negligible veiling.
Strong correlations between veiling and accretion diagnostics have been observed in CTTSs, including H$\alpha$ emission \citep{val93, Hartigan1995}, UV excess \citep{CalvetGullbring1998}, $\pab$ and $\brg$ emission \citep{muz98, nat04}, and the mass accretion rate $\macc$ \citep{CalvetGullbring1998, muz98, fio22dqtau, Nelissen2023}.

While magnetospheric accretion provides a well-established framework to explain veiling in CTTSs, continuum excess emission, and thus veiling, could in principle also be produced in boundary-layer accretion regimes through viscous dissipation at the disk--star interface \citep{LyndenBellPringle1974, Popham1995}. 
In boundary-layer accretion, the excess continuum emission arises from the braking and viscous dissipation of Keplerian disk material at the disk--star interface, rather than from localized accretion shocks. %As discussed by \citet{Hartmann1997}, this is expected to produce a more spatially extended emission region and a continuum spectrum that is relatively enhanced at optical and near-infrared wavelengths, rather than being strongly UV-dominated, 
As shown by \citet{Mendigutia2020} for more massive stars ($\gtrsim 2\,\msun$), the spatial and spectral distribution of BL emission differs from that expected in the MA scenario, and the resulting SED is more closely associated with a hot inner disk than with localized stellar hot spots. 
A clear comparison between magnetospheric hot-spot emission in CTTSs and disk-dominated emission in FU Orionis–type objects is presented by \citet{Hartmann2016}, highlighting how the continuum shape and wavelength dependence differ substantially between the two accretion regimes. 
These differences in the origin and temperature structure of the emitting region may therefore lead to veiling with spectral properties distinct from those observed in CTTSs dominated by MA \citep{Hartmann1997}.
%and differ in its wavelength dependence from the hot-spot emission characteristic of magnetospheric accretion \citep{Hartmann1997}, potentially leading to veiling with spectral properties different from those observed in CTTSs.

In Class\,I protostars, the measured veiling in the $K$ band ($r_K$) can reach values as high as $\sim 10$ \citep{Fiorellino2023}. It is likely that even higher values occur in very embedded sources for which veiling cannot be reliably measured. 
The $K$-band veiling has also been found to correlate with both the mid-IR (MIR) disk luminosity and the equivalent width of the $\brg$ line \citep[corrected for veiling --][]{Doppmann2003}. 
Part of this correlation may arise from the dependence of the dust sublimation radius on the total luminosity (stellar plus accretion); as the accretion luminosity increases, the dust sublimation radius moves outward, increasing the emitting surface area of the hot inner disk wall and thereby enhancing the NIR continuum excess \citep[e.g.,][]{Dullemond2001, Muzerolle2003, DAlessio2006}. 
Moreover, $r_K$ is systematically higher in protostars than in CTTSs, indicating a larger amount of circumstellar material and higher mass accretion rates in Class\,I sources \citep{dop05}. This supports the view that Class\,I and FS young stellar objects are actively accreting protostars, with accretion rates exceeding those typically observed in CTTSs.

\subsection{The CO bands} \label{sect:CObands}

Accreting YSOs show CO bands both in emission and absorption. CO band \textit{absorption} arises when cool molecular gas lies along the line of sight to a hotter background continuum source. In protostars, this absorption can originate in several locations, including the stellar photosphere (in late-type stars), the cool disk atmosphere (e.g., the outer disk or disk surface), and the inner envelope in Class\,I sources. 
The physical conditions required for CO absorption include ({\it i}) temperatures of $\sim 2{,}000$--$4{,}000$\,K, which are typically needed to significantly populate the vibrational levels responsible for strong band-head emission \citep[e.g.,][]{Najita1996}, ({\it ii}) a high CO column density, and ({\it iii}) a background continuum dominated by the stellar photosphere, hot dust emission, or the accretion continuum.
CO absorption occurs because the background continuum is bright, while the intervening CO gas absorbs photons in rovibrational transitions without emitting enough radiation to compensate.

CO absorption or emission depends on the mass accretion rate \citep{Calvet1991}.
CO band absorption is typically observed in Class\,II and Class\,I protostars, FU\,Orionis--like objects (hot disk midplan with the disk atmosphere absorbing), in systems with edge-on disks, and in sources with low or moderate accretion rates \citep{gre96, pontoppidan2005ices-411, Hartmann2016, connelley-reipurth2018}. 

\begin{figure}
    \centering
    \includegraphics[width=1\linewidth]{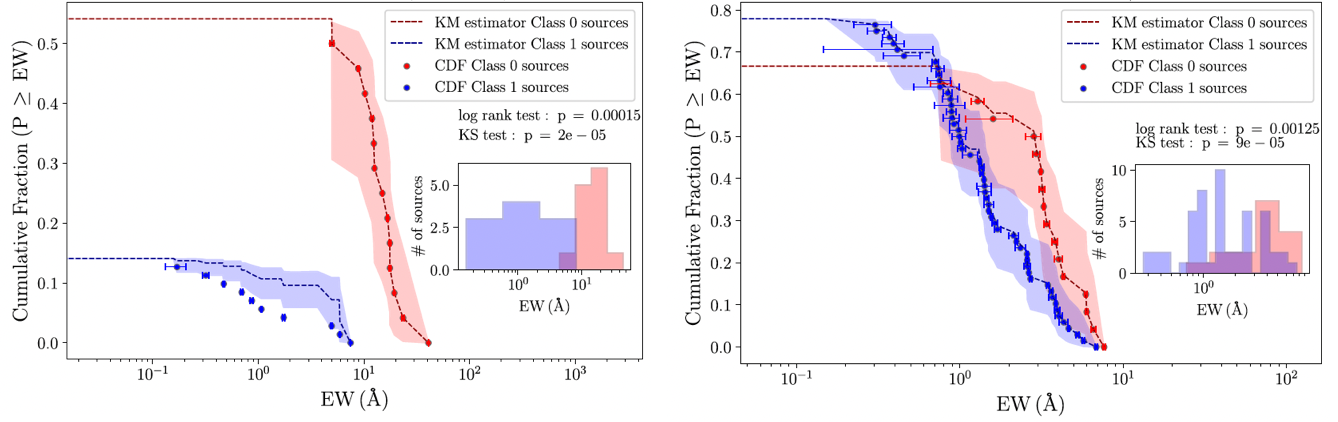}
    \caption{Comparisons of the CO ($v = 2 \rightarrow 1$) (left) and $\brg$ (right) emission-line between Class\,0 (red) and Class\,I (blue) protostars. The distribution of line parameter values is shown in the form of a CDF (normalized by the detection rate of the line in a given sample) and a histogram (shown in the sub-box on the right-hand side of the plot). Each point of the CDFs corresponds to one object’s measurement and uncertainty of the given line parameter. The p-value of the two-sided KS test is shown above the histograms on the right. The p-value is the probability that the two populations come from the same parent population. The resulting Kaplan–Meier estimator function for the  distributions with the dashed line and shaded area, respectively, which indicates the 95\% confidence interval of the survival function. In these cases, the resulting p-value of the log-rank statistical test, which takes into account non detections (unlike the KS test), is shown on the right. Adapted from \citet{LeGouellec2024}. © AAS. Reproduced with permission.}
    \label{fig:CO-BrG-Class0}
\end{figure}

CO band \textit{emission} originates in hot, dense molecular gas, typically located in the inner regions of circumstellar disks. It traces the inner gaseous disk at radii of $\lesssim 0.1$--$1$\,AU, inside or close to the dust sublimation radius, and is occasionally associated with dense disk winds. The typical physical conditions required for CO band emission are temperatures of $\sim 2{,}000$--$5{,}000$\,K and gas densities $\gtrsim 10^{10}$\,cm$^{-3}$.
When observed in emission, CO is thermally excited and the lines are optically thick or moderately optically thick. Emission occurs because the CO gas itself becomes an efficient emitter: it is sufficiently hot to populate high vibrational levels, collisional excitation is effective, and the emitted radiation exceeds the absorption of the background continuum.
CO band emission is typically observed in strongly accreting protostars, high-veiling systems, and sources with hot inner disks, and is often detected when the $\brg$ line is observed in emission \citep{dop05, con10}. CO emission is reported more frequently in Class\,0 (in about 60--70\% of sources) than in Class\,I protostars \citep[about 15\%,][]{Laos2021}, suggesting higher accretion activity in the former compared to the latter (\citealt{LeGouellec2024}, -- see Figure\,\ref{fig:CO-BrG-Class0}).

It is important to distinguish between the NIR CO overtone band-head emission at 2.3\micron\ discussed above and the CO fundamental emission at 4.6--5\micron. The latter is commonly detected towards deeply embedded Class\,0 protostars and traces warm molecular gas in the inner envelope–disk interface or in disk winds (as revealed by recent JWST observations -- \citealt{Federman2024, Rubinstein2024, vanDishoeck2025}). While both transitions probe warm gas, they arise under different excitation conditions and may trace distinct physical components of the embedded system.

Some young stars exhibit variability in their CO band profiles, which may be linked to changes in the accretion rate and the associated thermal and structural evolution of the inner disk \citep[e.g.,][]{Eisner2013, Bertelsen2016}. Variability can arise from accretion-driven changes in disk heating, modifications of the inner disk structure, variations in the continuum veiling level, or variations in disk geometry and inclination.
For instance, during strong accretion phases, the enhanced viscous heating of the disk interior and the irradiation of its surface can raise the temperature of the inner disk, promoting CO band emission. During quiescent phases, a cooler disk atmosphere overlying a hotter continuum-emitting region may instead produce CO bands in absorption. 
In contrast, the CO fundamental emission at 4.6–5\micron\ may remain in emission even when the overtone bands weaken or transition into absorption, as shown for deeply embedded sources such as I16253 \citep[e.g.,][]{Federman2024, Narang2026}. This difference likely reflects the distinct excitation conditions and emitting regions of the two transitions, with the fundamental one tracing warm molecular gas that is less directly tied to the hottest inner disk layers.

\subsection{Empirical relations as a tool to measure the accretion luminosity}
\label{sect:empirical-relations}

UVB, optical, and NIR H\,{\footnotesize I} emission lines, as well as lines from other species such as O\,{\footnotesize I}, He\,{\footnotesize I}, Ca\,{\footnotesize I}, Na\,{\footnotesize I}, and Fe\,{\footnotesize I}, are routinely found to exhibit empirical correlations with the accretion luminosity of CTTSs \citep[e.g.,][]{muz98, alc14, alc17, Fiorellino2025}. 
Notably, the luminosity of H\,{\footnotesize I} lines is found to correlate with the accretion luminosity even for transitions from high-\textit{n} atomic levels producing emission in the mid-infrared \citep{Salyk2013, Rigliaco2015, Komarova2020, Rogers2024, Testi2025, Tofflemire2025}, consistent with their origin in the hot, dense gas of magnetospheric accretion columns and associated shocks.
Thanks to JWST, emission from other species, such as CO, H$_2$O, OH, HCN, C$_2$H$_2$, and CO$_2$ have also been found to correlate with the accretion luminosity in the MIR \citep{Mallaney2026}, with a similar trend observed for the CO fundamental emission \citep{Rubinstein2024}.
However, some studies have shown that empirical relations based on MIR line diagnostics can be affected by contamination from disk winds or jet emission in some systems, potentially leading to an overestimate of the accretion luminosity if such contributions are not properly accounted for \citep[e.g.,][]{Bajaj2024}. 
Lastly, the empirical relation linking the Ly$\alpha$ flux to the integrated UV continuum between $91.2 - 165$\,nm found studying a sample of 16 Class\,II YSOs \citep{France2014} is at the basis of a recent attempt to constrain $\lacc$ on a Class\,0 protostar. 
This correlation has been proposed as a way to estimate the contribution of Ly$\alpha$ emission to the UV flux in the second absorption band of H$_2$O ($114–145$\,nm), which is responsible for the photodissociation of H$_2$O into OH.
%This correlation was used to estimate the contribution of Ly$\alpha$ emission to the UV flux in the second absorption band  of H$_2$O ($114–145$\,nm), which is responsible for the photodissociation of H$_2$O into OH \citep{Watson2025}.
In this Section, we review when and under which assumptions the diagnostics mentioned above have been applied to the analysis of protostars.

Empirical relations linking $\lacc$ to the luminosities of the Pa$\beta$ and Br$\gamma$ lines in CTTSs were first established by \citet{muz98}. In the same work, these relations were then applied to a sample of protostars in order to constrain their accretion luminosities. Prior to this study, the accretion luminosity of protostars was often taken to be comparable to their bolometric luminosity ($\lbol$), under the assumption that the intrinsic stellar luminosity was negligible.
Contrary to the expectations, using empirical relations calibrated on CTTS samples and assuming MA, \citet{muz98} and later \citet{Beck2007} found that the accretion luminosities of Class\,I sources were significantly lower than their bolometric luminosities and comparable to those of CTTSs. 
This result suggests that a substantial fraction of the disk material is accreted onto the forming star through episodic accretion events \citep{Fischer2023} or during earlier evolutionary stages, such as the Class\,0 phase.
Subsequent studies focusing on small samples of Class\,I protostars tested the $\lacc$--$L_{\rm Br\gamma}$ relation and found that it remains valid even for line luminosities up to an order of magnitude higher than those originally calibrated by \citet{muz98}. 
These works also showed that the majority of the protostars examined (five out of six) exhibits high accretion levels, with $\lacc$ reaching up to $\sim 50\%$ of the bolometric luminosity $\lbol$ \citep{nis05b, Greene2002}; however, the limited sample sizes prevented robust conclusions. 
More recent investigations have emphasized that the applicability of these empirical relations to deeply embedded protostars remains uncertain. In particular, radiative transfer effects, extinction, and scattering within the envelope may cause NIR tracers to probe only a fraction of the intrinsic magnetospheric emission.
For example, \citet{Harsono2023} showed that NIR accretion tracers may underestimate the accretion luminosity in protostars, as most of the magnetospheric emission can be completely obscured from view. In this scenario, the observed line fluxes trace only a small fraction of $\lacc$, arising primarily from scattered light along the less extincted cavity walls. As a consequence, more sophisticated approaches than the direct application of empirical relations are required for protostars in order to properly account for the occultation of the central source and the associated NIR emission \citep{Delabrosse2024, Fiorellino2025}.

\citet{Fiorellino2023} assembled a sample of 58 Class\,I and FS sources, including previously published observations. Using updated empirical relations \citep{alc17}, they estimated the accretion luminosity $\lacc$, stellar parameters, and the mass accretion rate ($\macc$) in a homogeneous and self-consistent manner across the entire sample \citep{ant08, Fiorellino2021}.
This approach relies on three main assumptions. ({\it i}) the total bolometric luminosity is assumed to be the sum of the intrinsic stellar luminosity and the accretion luminosity, with the disk emission treated as reprocessed accretion energy; ({\it ii}) the combined disk and envelope contributions, which cannot be disentangled in protostars, are parameterized through the $K$-band veiling; ({\it iii}) empirical relations calibrated on CTTSs between H\,{\footnotesize I} line luminosities and accretion luminosity are assumed to remain valid for Class\,I and FS sources.
Formally, ({\it i}) implies $\lbol = \lstar + \lacc$, where the disk luminosity ($L_{\rm disk}$) is included in $\lacc$; ({\it ii}) leads to the expression $M_{\rm bol} = BC_K + m_K + 2.5 \log (1+r_K) - A_K - 5\log (d/10\,{\rm pc})$,
where $r_K$ accounts for excess continuum emission; ({\it iii}) adopts the $\brg$ calibration
$\log \lacc = a \log L_{{\rm Br}\gamma} + b$, with $a = 1.19 \pm 0.10$ and $b = 4.02 \pm 0.51$ \citep{alc17}, supported by interferometric evidence that $\brg$ traces inner-disk accretion and ejection processes\footnote{VLTI/GRAVITY observations have confirmed that $\brg$ emission originates from accretion and ejection processes in the inner disk region \citep{Gravity2024Nowacki}.}.
Consequently, given the observed $K$-band magnitude ($m_K$), the $\brg$ luminosity ($L_{\rm Br\gamma}$), $\lbol$, the distance, and the $K$-band veiling ($r_K$), and assuming an age for the source (which sets the spectral type and thus the bolometric correction, $\mathrm{BC}_K$), the extinction ($A_K$) remains the only free parameter. Its value is determined by requiring consistency between the stellar luminosity $\lstar$ obtained from assumptions ({\it i}) and ({\it ii}). A detailed description of this method is provided in Section\,5.1 of \citet{Fiorellino2021}.

\begin{figure}
    \centering
    \includegraphics[width=0.45\linewidth]{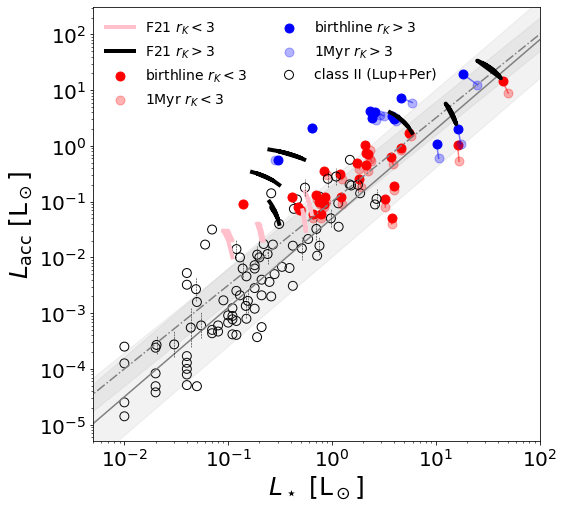}
    \includegraphics[width=0.45\linewidth]{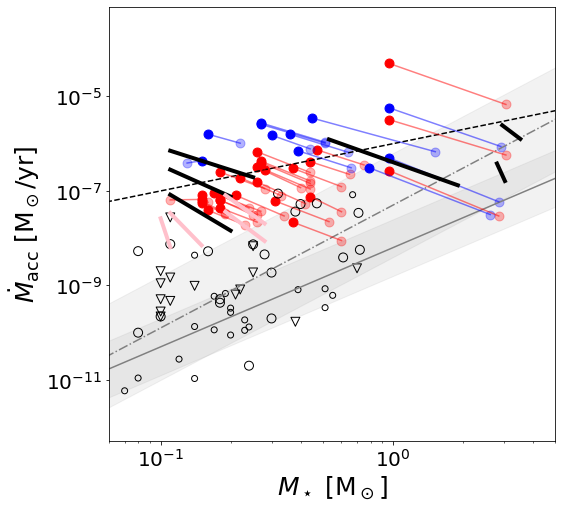}
    \caption{$\lacc - \lstar$ (left) and $\macc - \mstar$ (right) distributions showing accretion being more intense in Class\,I and FS than in Class\,II for sources with the same stellar luminosity or mass. Colored points correspond to Class\,I and FS as described in the legend, while empty circles correspond to Class\,II. Triangles are upper limits. 
    Adapted from \citet{Fiorellino2023}. © AAS. Reproduced with permission.}
    \label{fig:Fiorellino2023}
\end{figure}

In this way, \citet{Fiorellino2023} confirmed and generalized the main results previously obtained for the NGC~1333 cluster by \citet{Fiorellino2021}. Specifically, Class\,I and FS protostars exhibit accretion luminosities and mass accretion rates spanning the ranges $-2.5 < \log (\lacc/\lsun) < 0.5$ and $-8 < \log (\macc/\msun{\rm yr^{-1})} < -5$, respectively, in both cases systematically higher than those measured in CTTSs (see Figure\,\ref{fig:Fiorellino2023}). They also found that only a small fraction of sources (about 25\%) have accretion luminosities dominating the bolometric output, that is $\lacc/\lbol > 0.5$.
This latter result, however, is not confirmed by the subsequent study of \citet{Testi2025}, who analyzed a sample of Class\,I protostars in the MIR using empirical relations involving the Pf$\gamma$ and Br$\alpha$ lines derived in the same work. They found $\lacc/\lbol > 0.5$ for about 50\% of their sample, a fraction comparable to that reported for NGC\,1333 \citep[$\sim$40\%,][]{Fiorellino2021}. By contrast, using stellar masses measured from Keplerian disk rotation and comparing them with $L_{\rm bol}$, \citet{Hartmann2025} found that many protostars, including Class\,0 objects, exhibit low values of $\lacc/\lbol$. However, they caution that their conclusions may be affected by selection biases, as their sample is dominated by sources with large disks, and emphasize that larger and more homogeneous samples are required to draw firm conclusions on protostellar accretion and mass distributions.
Possible explanations for this tension are discussed in Section\,\ref{sect:lacc-discrepancy}.

The Br$\gamma$ line is also detected in Class\,0 protostars, with a frequency similar to that observed in Class\,I sources and exceeding 60\% \citep[][]{dop05, LeGouellec2024}. However, Class\,0 objects generally exhibit stronger Br$\gamma$ emission, with broader line widths (up to $\sim 200\,\mathrm{km\,s^{-1}}$), suggesting faster and more massive accretion flows \citep[][see also Figure\,\ref{fig:CO-BrG-Class0}]{Laos2021, LeGouellec2024}. More specifically, the velocity profiles appear to differ between the two protostellar classes: Class\,I sources often show blueshifted line centroids, whereas Class\,0 protostars tend to display symmetric profiles with little or no velocity shift. These differences have been suggested to indicate accretion mechanisms of different nature in Class\,0 systems \citep{LeGouellec2024}. Owing to the large uncertainties in the determination of $A_V$ and the possibility of different accretion geometries, measurements of the accretion luminosity in Class\,0 protostars are therefore even more challenging than in Class\,I sources. 

Given the severe observational limitations affecting accretion diagnostics in Class\,0 protostars, a number of indirect and highly model-dependent approaches have recently been explored.
This leads \citet{Watson2025} to suggest an alternative method based on OH spectral features produced by H$_2$O photodissociation. In this framework, the observed OH rotational ladder, once corrected for extinction, directly traces the rate at which H$_2$O molecules are dissociated by UV photons from the central source, since each dissociation event produces an excited OH molecule that cascades down the ladder.
The derived OH luminosity therefore provides a lower limit to the UV photon production rate in the 115–145\,nm band. The main source of uncertainty lies in estimating the fraction of UV photons absorbed by water, which depends on assumptions about the size and inflow geometry of the H$_2$O-emitting region.
After correcting for this effect, the inferred total UV flux is compared with predictions from magnetospheric accretion shock models, originally developed for CTTS systems, to derive the mass accretion rate.
Applying this method, the inferred mass accretion rate for the target Class\,0 protostar was $\dot M_{\rm acc} = (3.3 \pm 2.2) \times 10^{-10}\,M_\odot{\rm yr^{-1}}$, %similar than estimates obtained for the same source using alternative approaches \citep[$\macc = 5 \times 10^{-10}$,][]{Hsieh2017}, and 
well below typical accretion rates reported for protostars. 
%This indirect method exploits observational evidence from CTTSs to infer the UV flux required to photodissociate H$_2$O and reproduce the OH emission observed in the MIR. 
%This indirect method is based on three main assumptions: ({\it i}) the ratio between the UV flux and the IR continuum is the same for CTTSs and Class\,0 sources; ({\it ii}) the method used to compute $\lacc$ for Class\,0 protostars is the same as for Class\,II sources, i.e. accretion proceeds as described by the magnetospheric scenario; ({\it iii}) all the water has been photodissociated.
%Applying this method, the inferred mass accretion rate for the target Class\,0 protostar was $\dot M_{\rm acc} = (3.3 \pm 2.2) \times 10^{-10}\,M_\odot\,{\rm yr^{-1}}$, a value significantly  and well below typical accretion rates reported for protostars.
We point out that, as highlighted by the authors themselves, this approach has been applied to a single Class\,0 source showing spectral characteristics uncommon for this class. Indeed, the lack of detected water emission enables the assumption that most gas-phase H$_2$O is being photodissociated, while other classical accretion tracers are not clearly detected, likely owing to the high extinction typical of Class\,0 objects. 
Furthermore, this metodology relies on the assumption that MA regulates accretion in Class\,0 as it does in CTTSs, which has not yet been demonstrated. It therefore remains unclear whether this methodology can be straightforwardly extended to more embedded or less peculiar Class\,0 protostars.

\subsubsection{$\lacc - L_{HI}$
relationship: applicability and limitations}
\label{sect:empirical-rel-discussion}

Empirical relations linking emission line luminosities to accretion luminosity have become a widely employed tool to estimate mass accretion rates in young stellar objects. When applied within a self-consistent framework that simultaneously constrains extinction, veiling, and stellar properties, such relations provide an efficient way to characterize accretion across large samples. In particular, IR surveys have demonstrated the power of empirical relations to probe accretion in embedded sources that are otherwise inaccessible at optical wavelengths.

Nevertheless, a major source of uncertainty is represented by the high challenge of $A_V$ measurement, needed to deredden the H\,{\footnotesize I} fluxes to get corrected line luminosities. 
Moreover, these empirical relations are valid for, and calibrated on, CTTSs. As a consequence, their applicability to protostars requires careful consideration, given the physical differences between these evolutionary stages. \citet{Fiorellino2025} argue that the use of CTTS-based relations for Class\,I and FS sources is justified if these objects accrete through a magnetospheric process. Under this assumption, empirical relations involving H\,{\footnotesize I} recombination lines can be applied to Class\,I and FS sources to derive accretion luminosities. 
In support of this view, recent measurements suggest that magnetic field strengths in Class\,I protostars are comparable to those of CTTSs \citep{flores2024}, indicating that magnetospheric accretion may already operate at these stages. However, different observational techniques probe distinct components of the magnetic field. While \citet{flores2024} measure the total surface magnetic field strength through Zeeman broadening, thus being sensitive to both ordered and small-scale components, \citet{Drouglazet2026} employ spectropolarimetric observations that primarly trace the large-scale, organized magnetic field. 

In this context, \citet{Drouglazet2026} detect strong large-scale magnetic fields in only 6 out of the 15 Class\,I sources observed, with the remaining 9 objects showing no clear magnetic signature. 
These results suggest that, while strong magnetic fields may be present in Class\,I protostars, the large-scale field component required to efficiently truncate the disk and regulate magnetospheric accretion may not be equally developed in all sources, pointing to a diversity in magnetic field topology and accretion geometry within the Class\,I population. Furthermore, if Class\,I protostars host magnetic field strengths comparable to those of CTTSs but exhibit higher mass accretion rates, the balance between magnetic and accretion pressures is expected to shift, leading to smaller truncation radii. This effect should be taken into account when deriving $\macc$ from empirical relations calibrated on more evolved systems.

At the same time, important caveats must be kept in mind. Protostars are characterized by a much higher continuum veiling than CTTSs, with typical values reaching $r \sim 50$, compared to $r < 2$ in more evolved systems; as emphasized by \citet{Fiorellino2025}, neglecting or improperly accounting for veiling can lead to severe biases in line luminosities and, consequently, inferred accretion rates. 
In addition, in deeply embedded sources, part of the observed emission may arise from scattered or reprocessed radiation rather than directly tracing the accretion shock, further complicating the interpretation. 
These effects are likely to be particularly severe in Class\,0 protostars, for which magnetic fields are expected to be absent/weak and measurements are scarce, thus accretion may proceed through boundary-layer–like mechanisms rather than through magnetospheric funnel flows.

Empirical relations are therefore most reliable when applied to Class\,I and FS YSOs, provided that extinction and veiling are constrained and that the dominant accretion regime is magnetospheric. Outside these conditions, accretion luminosities inferred from empirical relations should be regarded as approximate or as lower limits.

\subsubsection{Implications for accretion onto planets} \label{sect:planet-accretion}

The use of empirical accretion relations has also motivated attempts to characterize accretion onto forming planets, where direct measurements of mass accretion rates are even more challenging. Observations of H$\alpha$ and NIR hydrogen emission from a small number of accreting protoplanets suggest that, at least phenomenologically, similar diagnostics may trace accretion-related processes \citep[e.g.,][]{Sallum2015, Haffert2019, Thanathibodee2019}. 

Part of the motivation for applying stellar accretion diagnostics to giant planets stems from the apparent continuity between the brown dwarf (BD) and planetary-mass regimes. Empirical accretion relations originally calibrated for CTTSs have been successfully extended to BDs, yielding accretion luminosities and mass accretion rates over several orders of magnitude in stellar mass \citep[e.g.,][]{Muzerolle2003, Natta2004, Herczeg2008, Almendros-Abad2024}. 
Since BDs have substantially lower masses and are expected to host weaker large-scale magnetic fields than CTTSs \citep[e.g.,][]{ReinersBasri2007, Morin2010}, their inclusion has further encouraged the exploratory extension of empirical accretion diagnostics toward even lower-mass objects.
Given that the mass boundary between brown dwarfs and giant planets is not sharply defined and may reflect formation history rather than a strict physical threshold \citep[e.g.,][]{Whitworth2007, Chabrier2014, Gilbert2025}, it is tempting to explore whether similar tools can be applied to giant forming planets.

However, the physical conditions of planetary accretion differ fundamentally from those of stellar and substellar accretion, including the geometry of the accretion flow, the depth of the gravitational potential, and the role of circumplanetary disks. 
Radiation hydrodynamic simulations indicate that gas accretes onto forming giant planets through a circumplanetary disk and forms a compact, high-temperature accretion shock at the planetary surface, with emission properties that depend sensitively on the shock structure and local opacity \citep[e.g.,][]{Szulagyi2020}. In this configuration, the geometry and thermodynamics of the accretion flow differ substantially from the magnetospheric paradigm commonly invoked for young stars.
As a consequence, accretion luminosities and mass accretion rates inferred for protoplanets obtained using $\lacc - \lline$ relationships remain highly uncertain and model-dependent, relying on assumptions about shock physics, emission efficiency, and radiative transfer \citep[e.g.,][]{Aoyama2019, Aoyama2021}. 

Nonetheless, constraining accretion onto planets is crucial for understanding the final masses of giant planets and for linking disk accretion during the protostellar phase to the early growth of planetary systems. Progress in this area will require both improved theoretical models of planetary accretion and high-sensitivity, high-angular-resolution observations capable of isolating planetary emission from the surrounding disk. In this context, improving our understanding of accretion mechanisms during the protostellar phase, when planet formation starts, is expected to provide valuable insights into planet formation, with direct implications for how accretion luminosities and mass accretion rates are inferred for forming planets.

\subsection{Outflows as a proxy of accretion} \label{sect:outflows}

Outflows and jets are a ubiquitous feature of the earliest stages of star formation and are commonly interpreted as indirect tracers of mass accretion in protostars \citep[e.g.,][]{Schwartz1977, Reipurth2001, Sicilia-Aguilar2020, Lee2020, Avachat2023}. 
In magneto-centrifugal launching scenarios, the ejection of mass and angular momentum through collimated jets and wider-angle molecular outflows is physically linked to disk-mediated accretion onto the central protostar, leading to an approximate proportionality between the mass-loss rate and the mass accretion rate \citep[e.g.,][]{Shu1994, Pudritz2007, Frank2014}. 
Observational support for this connection comes from correlations between accretion-related quantities and outflow properties, such as momentum flux and mass-loss rate $M_{\rm eje}$, measured through molecular tracers (such as CO) and atomic or ionic emission associated with jets \citep[e.g.,][]{Cabrit1992, Bontemps1996, Hartigan1995, Bally2016}. 

In deeply embedded protostars, where direct accretion diagnostics are often inaccessible because of high extinction and envelope opacity, outflows provide a valuable, though indirect, probe of accretion activity. [Fe\,{\footnotesize{II}}] and H$_2$ emission is routinely observed in both Class\,0 and Class\,I sources, with H$_2$ at 2.12\micron\ detected in emission in about 90--100\% of Class\,0 protostars and in only 40--50\% of Class\,I objects \citep{con10, Laos2021, LeGouellec2024, Federman2024, Narang2026}. 
However, the interpretation of outflow properties as instantaneous accretion tracers is not straightforward: molecular outflows integrate the accretion history over timescales of $10^{3}$--$10^{5}$\,yr, and episodic accretion can produce complex and non-linear signatures in the observed outflow energetics \citep[e.g.,][]{Vorobyov2006, Offner2011}. 
Indeed, different tracers probe distinct temporal baselines: while accretion tracers probe the instantaneous mass accretion rate, millimeter CO observations trace outflow activity on $\sim10^{4}$\,yr timescales, and far-IR CO emission samples shocks on intermediate ($\sim10^{2}$\,yr) timescales. Far-IR CO line luminosities from outflows have also been found to correlate with $L_{\rm bol}$, further supporting a statistical link between ejection and accretion in embedded sources \citep{Manoj2013, Manoj2016}.
This temporal integration is further compounded by the intrinsic variability of protostellar systems (see Section\,\ref{sect:variability}), which decouples short-term accretion fluctuations from the large-scale outflow response. Consequently, while outflows are robust indicators of ongoing or recent accretion in protostars, they primarily trace time-averaged accretion rates rather than instantaneous values. 
Consistently, using $L_{\rm bol}$ as a proxy for $L_{\rm acc}$ and combining it with [Fe\,{\footnotesize{II}}] and [O\,{\footnotesize{I}}] diagnostics of jets, \citet{Watson2016} estimated that the mass ejection rate in Class\,0 protostars corresponds to $\sim$10\% of the mass accretion rate, in agreement with expectations from magneto-centrifugal launching models.
Despite these caveats, recent observational studies have provided further empirical support for a close connection between accretion and ejection in embedded protostars. In particular, \citet{Ghosh2026} found that the empirical correlations between the 3.3\,mm continuum flux and the Br$\gamma$ line flux, as well as between the mass accretion rate and the ionized mass-loss rate previously established for Class\,II sources, also hold for Class\,I protostars. 
Recent ALMA studies have begun a quantitative exploration of the connection between $\macc$ and $\dot{M}_{\rm eje}$ on a statistical basis by deriving accretion rates directly from jet properties. For instance, \citet{Dutta2024} estimated jet mass-loss rates from CO emission associated with high-velocity knots and inferred accretion rates by combining the bolometric and jet kinetic luminosities under the assumption that both are powered by accretion. In this framework, they obtained $\dot{M}_{\rm eje}/\macc$ ratios spanning $\sim 0.003$ to $\sim 2$, significantly broader than the typically used value of $\sim 0.1$ \citep{Hartigan1995}.
These estimates, however, are affected by substantial uncertainties; the derivation of $\dot{M}_{\rm eje}$ depends on assumptions on excitation temperature, CO abundance, geometry, and inclination, and is likely to provide a lower limit, as it primarily probes the dense jet axis rather than the full outflow. Moreover, the inferred accretion rates rely on simplified assumptions on the stellar mass and radius, and on the fraction of $\lbol$ attributable to accretion. 
Overall, these factors can introduce uncertainties of at least tens of percent and potentially larger systematic biases.
Despite these limitations, such studies demonstrate that, at least in a statistical sense, outflow properties retain a measurable link to accretion even in deeply embedded protostars.

High-angular-resolution observations have also revealed that protostellar jets and outflows are not only powerful but also highly collimated, even at distances smaller than $\sim 100$\,au from the central protostar \citep[e.g.,][]{Bjerkeli2016, Lee2018}. 
JWST observations are providing unprecedented insight into jets from deeply embedded protostars, revealing both their molecular and atomic structure at high spatial resolution. For instance, studies of Class\,0 sources such as HH211 and IRAS 15398-3359 have shown highly collimated jets with layered molecular and atomic components, while recent IFU observations have resolved the kinematics of jets in sources such as B335 and HOPS 153 (e.g., \citealt{Yang2022, Narang2025, Federman2026}). Similarly, ALMA observations of protostellar systems have shown that molecular and atomic jets maintain narrow opening angles close to their launching regions, supporting magneto-centrifugal models in which jets efficiently extract angular momentum from the disk and thereby regulate accretion onto the protostar \citep[e.g.,][]{Shang1998, Lee2017, Hirota2017, Tabone2017, Bally2023}. 
This early collimation implies that jets can interact with the surrounding envelope and natal core in a highly directional manner, carving cavities, entraining molecular gas, and injecting momentum and energy into the immediate environment \citep{Shu1991, Mendoza2004, Shang2006, Liang2020}. Such feedback can influence both the infall geometry and the efficiency of mass accretion, as well as the fragmentation and turbulence of the parent core \citep[e.g.,][]{Arce2010, Offner2014}. Numerical simulations further indicate that collimated outflows play a crucial role in regulating star formation by limiting stellar masses, removing excess angular momentum, and driving turbulence on core and clump scales \citep[e.g.,][]{Federrath2014, Federrath2015}. Together, these observational and theoretical results suggest that outflow collimation is a key ingredient in the self-regulation of accretion during the embedded phases, linking small-scale jet-launching physics to larger-scale feedback processes in star-forming regions.

\subsection{Protostellar disk and envelope mass and size}
\label{sect:disk-envelope-mass}

The disk and envelope contributions are fundamental when studying protostellar systems. In particular, the accretion process directly links the flow of material from the disk into the forming star, and recently discovered streamers may play a role in the accretion process as well. 
The formation and evolution of these structures are governed by the redistribution of angular momentum during collapse and are strongly influenced by magnetic fields, which regulate disk formation, braking, and mass transport.
In this Section, we review methodologies to constrain the disk and envelope properties, such as their mass, radius, and possible sub-structures, as well as how they are linked to the accretion process.

High-angular-resolution interferometric observations with facilities such as ALMA, NOEMA\footnote{NOrthern Extended Millimeter Array}, and the VLA\footnote{Very Large Array} have revolutionized the study of protostellar disks and envelopes, enabling direct constraints on disk kinematics and mass distributions at spatial scales of a few tens of astronomical units. 
Disk masses in protostellar systems are most commonly estimated from (sub-)millimeter continuum observations, under the assumption that the emission is dominated by optically thin dust and can be converted into a total disk mass by adopting a dust opacity, a characteristic dust temperature, and a gas-to-dust mass ratio \citep[e.g.,][]{Hildebrand1983, Beckwith1990}. While this approach has proven effective for Class\,II disks (albeit with caveats - see \citealt{Miotello2023PPVII}), its application to embedded protostars is subject to significant uncertainties. First of all, dust emission from the surrounding envelope can contaminate the disk flux; furthermore, parts of the disk may be optically thick even at millimeter wavelengths, and the dust temperature and opacity are poorly constrained in the presence of strong irradiation and accretion heating \citep{Jorgensen2009, Dunham2014, Segura-Cox2018, Maury2019}. Envelope masses are similarly derived from continuum or molecular line observations, but disentangling envelope and disk contributions requires high-angular-resolution data and often relies on parametric modeling of the density and temperature structure, introducing additional degeneracies.

To mitigate these limitations, more recent studies have combined continuum observations with kinematic information from molecular lines to isolate the rotationally supported disk component and to constrain disk radii ($R_{\rm disk}$) and masses ($M_{\rm disk}$) more robustly. The resulting measurements have mean values of $R_{\rm disk} \sim 30$--$50$\,au and $M_{\rm disk} \sim 10$--$30$\,M$_{\oplus}$, depending on the evolutionary stage. However, the evolutionary trend of disk properties remains debated; while some studies suggest that disk radii and masses decrease from Class\,0 to Class\,I and FS sources \citep[e.g.,][]{Tobin2020}, other works employing radiative transfer models with variable dust properties do not find strong evidence for a systematic decline in disk mass along the embedded phases \citep[e.g.,][]{Sheehan2022}. 
Radiative transfer modeling of both dust and gas emission has also been employed to account for optical depth effects and temperature gradients within the disk and inner envelope. 
Nevertheless, while disk radii estimates have been demonstrated to be accurate and reliable, significant systematic uncertainties remain on disk masses when compared to post-processing results (see Section\,\ref{subsubsec:synthetic_obs} for a dedicated discussion). 
$M_{\rm disk}$ estimates in protostars can vary by factors of a few depending on the adopted assumptions on dust properties, temperature structure, and the treatment of envelope contamination. 
\citet{Tung2024} showed that optical depth effects, dust evolution, and unresolved substructures can lead to substantial underestimates of the true disk mass when relying on simple continuum-based methods. 
These results highlight that, while the constraints on disk masses are reaching increasing spatial and kinematic detail, their absolute values in embedded protostars remain intrinsically uncertain.

\begin{figure}
    \centering
    \includegraphics[width=0.45\linewidth]{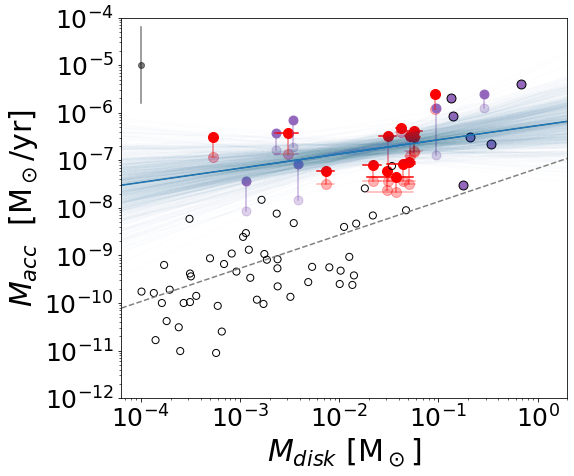}
    
    \caption{Mass accretion rate vs. disk dust mass. 
    Red and purple filled circles are the Class\,I sources. 
    Empty circles are Class\,II from Lupus. The blue line corresponds to the best fit for the overall sample of Class\,I YSOs, while the light blue lines are a subsample of the results of some chains. The dashed grey shows the best fit for Class\,II YSOs \citep[][]{Manara2016-evidenceacc}. Adapted from \citet{Fiorellino2022b}. © AAS. Reproduced with permission.}
    \label{fig:Fiorellino2022}
\end{figure}

Taking in mind these uncertainties, recent studies have begun to establish empirical links between disk properties and accretion in embedded systems. \citet{Fiorellino2022b} found a statistically significant correlation between the mass accretion rate and the disk mass in Class\,I protostars -- displayed in Figure\,\ref{fig:Fiorellino2022}, extending to earlier evolutionary stages the \(\dot{M}_{\rm acc}\)--\(M_{\rm disk}\) relation previously identified in Class\,II disks \citep{Manara2016-evidenceacc}. 
A linear correlation between the logaritm of these properties is expected to naturally arise in Class\,II objects in the viscous magnetospheric accretion framework \citep{Lodato2017}; although the $\log \dot{M}_{\rm acc} - \log M_{\rm disk}$ correlation in Class\,I sources appears flatter and exhibits larger scatter, the presence of such a correlation suggests that the disk mechanisms such as viscosity might already play a role in regulating accretion during the embedded phase. 
At the same time, theoretical and observational considerations indicate that in Class\,0/I systems the relevant mass reservoir may not be limited to the disk alone. \citet{Mendigutia2018} argued that a correlation between $\macc$ and the total circumstellar mass, including both disk and envelope ($M_{\rm disk}+M_{\rm env}$), is expected during the embedded phases, reflecting the ongoing interaction between the forming star, its disk, and the surrounding envelope. In this picture, accretion rates are not regulated solely by internal disk viscous evolution, but are also influenced by large-scale mass infall and environmental feeding.
A complete analysis of this relation should be performed with models also considering the combined effects of envelope replenishment, disk growth, and time-variable accretion. 

Beyond the disk itself, high-sensitivity molecular line observations have uncovered increasingly complex kinematics in protostellar envelopes. Several studies have identified large-scale infalling structures, often referred to as accretion streamers, connecting the surrounding molecular core to the disk \citep[e.g.,][]{tokuda2018warm-1ee, Pineda2020, Garufi2021, gupta2023reflections-8e0, kuffmeier2023rejuvenating-73b, ValdiviaMena2024}. These asymmetric inflow patterns challenge the classical picture of spherical or axisymmetric collapse and indicate that disk growth may be governed by highly structured, time-dependent mass delivery from the envelope (as also found in large scale numerical simulations -- see Section \ref{subsubsec:largescale}). While the detailed connection between these large-scale streamers and the accretion processes operating in the inner disk remains an open question, their presence suggests that envelope dynamics play an important role in setting the boundary conditions for disk accretion.

Chemical diagnostics provide an additional tool to probe the disk--envelope interface. Specific molecular tracers, such as N$_2$H$^+$ and C$^{18}$O, are commonly used to isolate dense gas and to identify chemical differentiation across the transition region between the infalling envelope and the forming disk \citep[e.g.,][]{Oberg2011, Harsono2015, vanTerwisga2019}. Spatial variations in molecular abundances and excitation conditions offer complementary constraints on the physical structure and kinematics of the accreting material, helping to disentangle rotationally supported disk gas from infalling envelope components. 

\subsection{The variability of protostars}
\label{sect:variability}

Protostellar accretion is increasingly recognized as a highly variable process, yet its temporal behavior remains poorly constrained, particularly during the deeply embedded phases. The main reason is the embedded nature of protostars, which prevents systematic monitoring at optical wavelengths; as a consequence, studies of protostellar variability - and more specifically of $\lacc$ variability - rely almost entirely on infrared time-domain surveys.

Mid-infrared (MIR) monitoring campaigns have provided the first systematic view of protostellar variability on timescales from months to years. In particular, time-domain observations from WISE\footnote{Wide-field Infrared Survey Explorer} and NEOWISE\footnote{Near-Earth Object Wide-field Infrared Survey Explorer} have widespread MIR variability among Class\,0 and Class\,I protostars, with light curves displaying both stochastic fluctuations and long-term trends \citep[e.g.,][]{Morales-Calderon2011, Fischer2019, Park2021, Zakri2022, Chinmay2026}. 
Because MIR emission traces warm dust in the inner disk and envelope, such variability is commonly interpreted as reflecting changes in the accretion luminosity, disk heating, or inner disk geometry \citep{ContrerasPena2020}. However, disentangling these contributions remains challenging, limiting a direct interpretation of MIR variability in terms of instantaneous accretion rates.
NIR time-domain surveys have further advanced the study of protostellar variability by probing shorter wavelengths while still mitigating the effects of extinction. In particular, the ESO Public Survey Vista Variables in the Vía Láctea (VVV) has provided multi-epoch $K_s$-band light curves for millions of sources across the Galactic bulge and inner disk \citep[e.g.,][]{Minniti2010, Saito2012}. 
Despite these advances, NIR monitoring programs specifically designed to characterize non-episodic protostellar accretion variability, to compare protostellar and PMS variability, remain scarce.

More extreme manifestations of protostellar variability are provided by episodic accretion events, such as FU\,Orionis- and EX\,Lupi-type outbursts, during which the accretion rate increases by several orders of magnitude for periods ranging from months to decades. 
Infrared and submillimeter spectroscopy during these events has revealed dramatic changes in disk structure, chemistry, and thermal balance, supporting the idea that episodic accretion plays a major role in the mass assembly of young stars \citep[e.g.,][]{Hartmann1996, Audard2014}. 
Growing observational evidence suggests that such bursts are not rare events, but instead represent a recurrent phase of protostellar evolution, consistent with predictions from models invoking disk gravitational instability and envelope-fed accretion \citep[e.g.,][]{Vorobyov2015, Fischer2023}. 
VVV data have enabled systematic searches for variable and eruptive young stellar objects, revealing numerous YSOs - predominantly Class\,I systems - with variability spanning from days to years \citep[e.g.,][]{Guo2020}. 
Combined spectroscopic and photometric analyses show that extremely strong outbursts (FUors–like) are characterized by distinctive NIR spectral changes correlated with luminosity increases, consistent with enhanced accretion activity. 
The effort by \citet{Contreras-Pena2025} to systematically collect and classify young stars undergoing episodic accretion further indicates that a substantial fraction of embedded protostars may experience accretion bursts that remain undetected at optical wavelengths due to extinction, emphasizing the importance of infrared monitoring.

Quantifying the incidence of episodic accretion during the protostellar phase remains challenging, primarily because accretion bursts are intrinsically rare, transient, and often obscured. Nevertheless, recent infrared and submillimeter time-domain surveys have begun to place statistical constraints on how frequently protostars undergo accretion outbursts. Near-infrared monitoring of large samples of embedded young stellar objects with the VVV and VVVX\footnote{VISTA Variables in the Vía Láctea eXtended} surveys indicates that approximately 2--3\% of Class\,I protostars exhibit high-amplitude ($\Delta K_s \gtrsim 2$ mag) variability consistent with episodic accretion at any given time \citep{Contreras-Pena2024}. After accounting for survey completeness and temporal sampling, these results imply recurrence timescales of order $\sim10^{3}$--$10^{4}$\,yr for major, FU\,Orionis--like outbursts during the Class\,I phase \citep{Fischer2019}. Complementary constraints from MIR and submillimeter monitoring surveys, including \textit{Spitzer}, WISE/NEOWISE, and the JCMT Transient Survey, suggest that lower-amplitude accretion variability is more common, with a substantial fraction ($\gtrsim10$--20\%) of protostars exhibiting measurable luminosity changes on timescales of months to years \citep[e.g.,][]{Johnstone2018, Fischer2019, Lee2021}. 
While robust incidence estimates for Class\,0 and FS sources remain scarce, these observations indicate that episodic accretion is a recurrent, though temporally sparse, component of protostellar mass assembly rather than an exceptional phenomenon.

The prevalence of strong accretion variability during the protostellar phase has important implications for early disk evolution and planet formation. On one hand, the large mass reservoirs available in young disks suggest that planet formation could begin already during the embedded stages \citep{tyc20}. 
On the other hand, the highly dynamic accretion environment, characterized by frequent luminosity bursts and rapid structural disk changes, raises questions about the survivability and growth of planet-forming solids \citep[e.g.,][]{Vorobyov2006, Audard2014, Cieza2016, kadam2022primordial-953, das2025accretion-8d9}. 
If disk instabilities and episodic accretion are as common as current observations suggest, it remains unclear how these interact with the {\it standard} magnetospheric accretion occurring during quiescence and how efficiently planets can form and evolve under such conditions. Understanding the interplay between accretion variability, disk evolution, and early planet formation therefore represents one of the major open questions in the study of protostellar accretion.

\section{Models and simulations}\label{sec:models_and_sims}

Numerical simulations of star formation provide a valid tool to explore the influence of physical mechanisms, initial conditions, and environments on the assembly and evolution of protostars. Since the pioneering mono-dimensional studies of \cite{larson1969numerical-ed8}, \cite{penston1969dynamics-054}, and \cite{shu77}, the field has made tremendous progress: modern simulations in both two- and three-dimensional configuration explore the effects of hydrodynamics, radiative transfer, and magnetic fields both in the ideal and non-ideal approximation, and provide diagnostics for evolutionary and accretion parameter to compare against the observational constraints. In this Section, we aim at summarizing the numerical effort carried out in the last few decades: we give an overview of the three main computational techniques in Section \ref{subsec:numerical_overview}, and discuss the simulations and their results in \ref{subsec:pure_hydro},\ref{subsec:rad_hydro}, and \ref{subsec:MHD}, following increasing physical complexity. %{\textcolor{blue}{Non ho scritto da nessuna parte che il disco è una soluzione al problema dell'angular momentum. Secondo te manca?} mi sembra di averlo scritto io da qualche parte? se non lo abbiamo scritto, dobbiamo aggiungerlo.}

\subsection{Numerical methods: an overview}\label{subsec:numerical_overview}

Simulations of cloud/core collapse leading to protoplanetary disk formation can be performed with three main numerical techniques: thin disk approximation, Smoothed Particle Hydrodynamics (SPH), and Adaptive Mesh Refinement (AMR). In the following subsections, we briefly present each of these approaches and discuss their strengths and limitations.

\subsubsection{Thin disk (two-dimensional models)}

Thin-disk simulations assume that vertical force balance is established on timescales shorter than those governing cloud contraction. Isothermal, non-rotating, magnetized cylindrical or spherical clouds quickly reach vertical hydrostatic equilibrium - fast enough to beat the contraction in the direction perpendicular to the field lines caused by ambipolar diffusion \citep{fiedler1993ambipolar-b51}. This motivates the choice of a geometrically thin disk, with a scale height smaller than its radial extent, as model cloud. This magnetized thin disk approximation, originally proposed by \cite{ciolek1993ambipolar-0f6} and \cite{basu1994magnetic-176}, is one of the simplest numerical frameworks to address the problem of a contracting cloud; the vertical integration reduces the dimensions to two, thereby resulting in an inexpensive setup that can be evolved up to the virtually entire Class I lifetime (of the order of $\sim$ Myr). This perk comes at the expense of a realistic description of the the magnetic field geometry, as well as the disk vertical structure; the neglect of vertical stratification and consequently three-dimensional instabilities may lead to different fragmentation thresholds compared to those derived in full 3D simulations.

%has been extensively employed in the modeling of collapsing protostellar cores (see, e.g., \citealt{vorobyov2005effect-a0d, vorobyov2006burst-f6a, vorobyov2010burst-a5d, vorobyov2015variable-98b, vorobyov2018early-117, vorobyov2020accretion-65a, bae2014accretion-224})

%Vorobyov et al 2018: In this paper, we continue our efforts to study the long-term evolution of self-gravitating and viscous circumstellar disks started in Vorobyov & Basu (2009). We now employ a numerical hydrodynamics code that in addition to the gaseous component also includes the dust component. 

\subsubsection{Smoothed Particles Hydrodynamics (SPH)}

Smoothed Particles Hydrodynamics (SPH), originally formulated by \cite{lucy1977numerical-698} and \cite{gingold1977smoothed-263}, is a Lagrangian scheme to solve the equations of hydrodynamics by approximating the continuum dynamics of fluids through the use of particles with properties smoothed over a local kernel. Lagrangian schemes are a class of particle-based numerical frameworks, as opposed to Eulerian schemes, which instead are grid-based; Eulerian codes feature a uniform, fixed Cartesian mesh (i.e., with cells aligned to the axes of a Cartesian coordinate system), while Lagrangian schemes employ a moving mesh that follows the flow of the fluid. This characteristic makes Lagrangian methods intrinsically adaptive, in that changes in the density and flow morphology are accounted for without the need for mesh refinement. This also implies that resolution is automatically concentrated in regions of high particle density, avoiding waste of computational resources outside the regions of interest. Another major perk of the Lagrangian SPH framework is the built-in conservation properties in the equation of motion: linear momentum, angular momentum, and entropy are simultaneously exactly conserved independently of configuration, which makes SPH particularly well suited to treat complex (3D) geometries. 
The main limitation of SPH simulations is the lack of an explicit treatment of physical viscosity, which is addressed by the addition of an artificial viscosity to capture shocks and dissipative processes. 
Such artificial viscosity can be implemented in multiple ways, which in turn can impact the treatment of angular momentum transport; however, most artificial viscosities can be directly translated into a combination of Navier-Stokes shear and bulk viscosity terms and can therefore be built in a physically-motivated way (see, e.g., \citealt{flebbe1994smoothed-0ad, watkins1996new-07d}). For a review on SPH methods and their astrophysical approaches, see \cite{monaghan1992smoothed-c47, monaghan2005smoothed-e50, Price2004, rosswog2009astrophysical-6bd, springel2010smoothed-312}.

\subsubsection{Adaptive Mesh Refinement (AMR)}

The Adaptive Mesh Refinement (AMR) method, originally proposed by \cite{berger1984adaptive-3e5} and \cite{berger1989local-788}, is a Eulerian scheme which covers the computational domain by a hierarchy of nested grids that are dynamically refined in regions where higher resolution is required. The main advantage of AMR is its flexibility, in that it simultaneously resolves larger and smaller scales; furthermore, in contrast to SPH methods, shocks and discontinuities are easily captured without the need for artificial terms, and the Eulerian grid allows for a straightforward implementation of non-ideal MHD terms, radiative diffusion, and chemical cooling. The price to pay for such detailed simulations is a substantial computational cost, which limits the evolutionary timescales that can be probed with AMR codes. For core collapse simulations, this usually leads to an evolution up to a maximum of a few $10^5$ yr. Furthermore, to prevent the timestep to become prohibitively small near the central object, protostars are usually replaced by \textit{sink particles} - unresolved particles that can accrete and interact with the other particles in the simulation only through gravity. The criteria to transform an over dense gas particle and its neighbors into a sink particle involve not only density, but also the smoothing length, as well as tests to determine whether the particles may be in the process of tidal disruption or bouncing, ensuring that sinks only replace groups of particles that would otherwise continue to collapse (see \citealt{bate1995modelling-624} for details). The use of sink particles, while extremely functional to keep the computational time under control and allow to proceed past the formation of the first protostars, has the disadvantage of introducing unresolved regions; this impacts the treatment of internal processes, such as the emission of radiation from the accreting gas and radiation feedback, and does not allow to separate the inner disk from the protostar itself, which influences the determination of accretion properties.

\subsubsection{Summary}

The three numerical approaches described above - thin disk approximation, SPH, and AMR - provide a complementary perspective to the problem of cloud collapse simulations. While none of the three is able to simultaneously ({\it i}) resolve the large-scale collapse, ({\it ii}) capture the detailed physics of accretion onto the protostars, and ({\it iii}) include radiative feedback processes, they all play a crucial role depending on the nature of the science question. \textit{Thin disk} models are the optimal choice to study the long-term evolution of protostars, virtually up to the end of the protoplanetary disk (Class II) phase, while they do not allow the detailed modeling of accretion and radiative feedback; \textit{SPH} is a powerful tool for three dimensional simulations, with its adaptive resolution and intrinsic conservation properties, but the presence of an artificial viscosity limits the resolution of shocks and the treatment of radiation and magnetic fields; finally, \textit{AMR} is excellent at resolving the multi-scale nature of the collapse and capturing the details of non-ideal MHD effects, as well as radiation transport and dynamical feedback, but its extreme computational cost restricts its use to shorter timescales and smaller samples. The choice of the numerical approach is therefore strongly dependent on the goal of the simulation; the current state of the art is a combination of works that have employed all three methods, as we discuss in the following.

%\cite{vorobyov2005effect-a0d, vorobyov2010burst-a5d, vorobyov2015variable-98b, vorobyov2020accretion-65a}: seminal grid-based, thin-disc calculations. Focus on luminosity bursts (FUor eruptions). Believed that FUor bursts are caused by sharp increases in the mass accretion from the disk onto the star. Can be explained with disk instability or perturbations to the disk, driven by external or internal agents (tidal effects due to the presence of a companion, \citealt{bonnell1992binary-9e0}; infall of mass from the outer disk at a rate sufficient to ionise hydrogen in the inner disk, \citealt{bell1993using-c37, riaz2018episodic-f5f}; magnetically layered structure, \citealt{armitage2001episodic-7c9}; massive planets embedded in the disk, \citealt{lodato2004massive-6a8}; infall of clumps originated by disk instability, \citealt{vorobyov2005origin-b16}; encounters of star-disk systems in dense clusters, \citealt{pfalzner2008accretion-d11}; combinations of MRI and GI, \citealt{zhu2009nonsteady-013}; star-star encounters, \citealt{forgan2010stellar-782}; infall of protoplanetary mass objects formed within the disk, \citealt{machida2011recurrent-218}; tidal disruption of young, massive planets near their host star, \citealt{nayakshin2012fu-57f}; non-zero residual viscosity in dead zones, \citealt{bae2014accretion-224}; environmental factors \citealt{kuffmeier2018episodic-0c6})

\subsection{Pure hydrodynamic simulations}\label{subsec:pure_hydro}

The first generation of cloud collapse simulations featured a pure hydrodynamical approach with self-gravity, typically employing the thin disk approximation or SPH.

\subsubsection{The Vorobyov $\&$ Basu series: a historical perspective}

\begin{figure*}
    \centering
    \includegraphics[width=\linewidth]{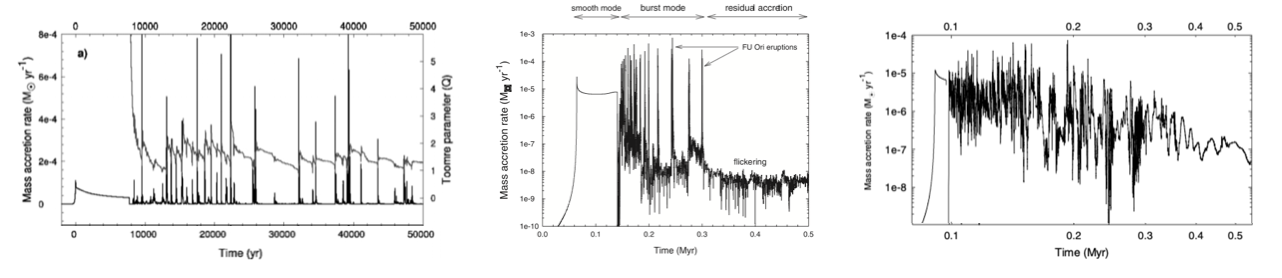}
    \caption{Time evolution of the accretion rate in thin-disc approximation simulations. Adapted from \cite{vorobyov2005origin-9cd, vorobyov2006burst-b61, vorobyov2010burst-a5d} (panels from left to right). © AAS. Reproduced with permission.}
    \label{fig:VB050610}
\end{figure*}

Vorobyov $\&$ Basu were the first to couple the collapse of a rotating pre-stellar core to the formation of protostars and accretion disks around them, leading to a self-consistent accretion history. Their seminal series of papers presented suites of simulations in the thin disk approximation, starting from the collapse of an originally starless, magnetically supercritical core (i.e., with the effect of the magnetic field comparable to, but weaker than, that of gravity).
%The original, simple model (presented in \citealt{vorobyov2005effect-aef}) was then expanded and gradually complicated to include additional physical effects; from a detailed energy balance equation accounting for radiative cooling, viscous and shock heating, and heating due to the stellar and background irradiation \cite{vorobyov2010burst-a5d, vorobyov2015variable-98b}, to dust in the two-population approximation of \citep{Birnstiel2012, vorobyov2018early-117} and metallicity \cite{vorobyov2020accretion-65a}. 

The early papers \citep{vorobyov2005effect-aef, vorobyov2005origin-9cd, vorobyov2006burst-b61} focused on the onset of instability in the star formation process, showing how the protostar and protoplanetary disk, formed as a result of the collapse of a single core, quickly become unstable to the formation of a spiral, which originates from the continuous infall of material from the envelope. Within the spiral arms, fragmentation takes place as dense protoplanetary clumps come together and occasionally fall onto the protostar, leading to episodes of vigorous accretion (as high as $10^{-4}$ M$_{\odot}$yr$^{-1}$) as opposed to the ``quiescent" periods (where the accretion rate is of the order of  $10^{-7}$ -- $10^{-6}$ M$_{\odot}$yr$^{-1}$). Figure \ref{fig:VB050610} shows an example of the highly variable accretion rates resulting from these simulations.
Subsequent works \citep{vorobyov2007selfregulated-8f3, vorobyov2008mass-38b} have further evolved the disks well into the Class II stage, following their accretion history as shaped by the gravitational torques produced by low amplitude, non-axisymmetric density perturbations that replace the early spiral structure. 
The original model was later expanded and gradually complicated to include additional effects, from improved thermal physics introducing a detailed energy balance equation accounting for radiative cooling, viscous and shock heating, as well as heating due to the stellar and background irradiation \citep{vorobyov2010burst-a5d, vorobyov2015variable-98b}, to dust (in the two-population approximation of \citealt{Birnstiel2012} -- \citealt{vorobyov2018early-117}) and metallicity \cite{vorobyov2020accretion-65a}. 

This series of thin-disk models of collapsing cores played a major role in understanding the accretion processes in the protostellar phase, tackling both the high time variability and the accretion outbursts dominating the mass budget in the Class 0/I phase. %\textcolor{red}{Doug: might be worth noting here the physics that is not included - and therefore the caveats that should be kept in mind when using these models as a proxy for how real systems evolve}

\subsubsection{Early SPH and grid-based simulations}

\begin{figure*}
    \centering
    \includegraphics[width=0.95\linewidth]{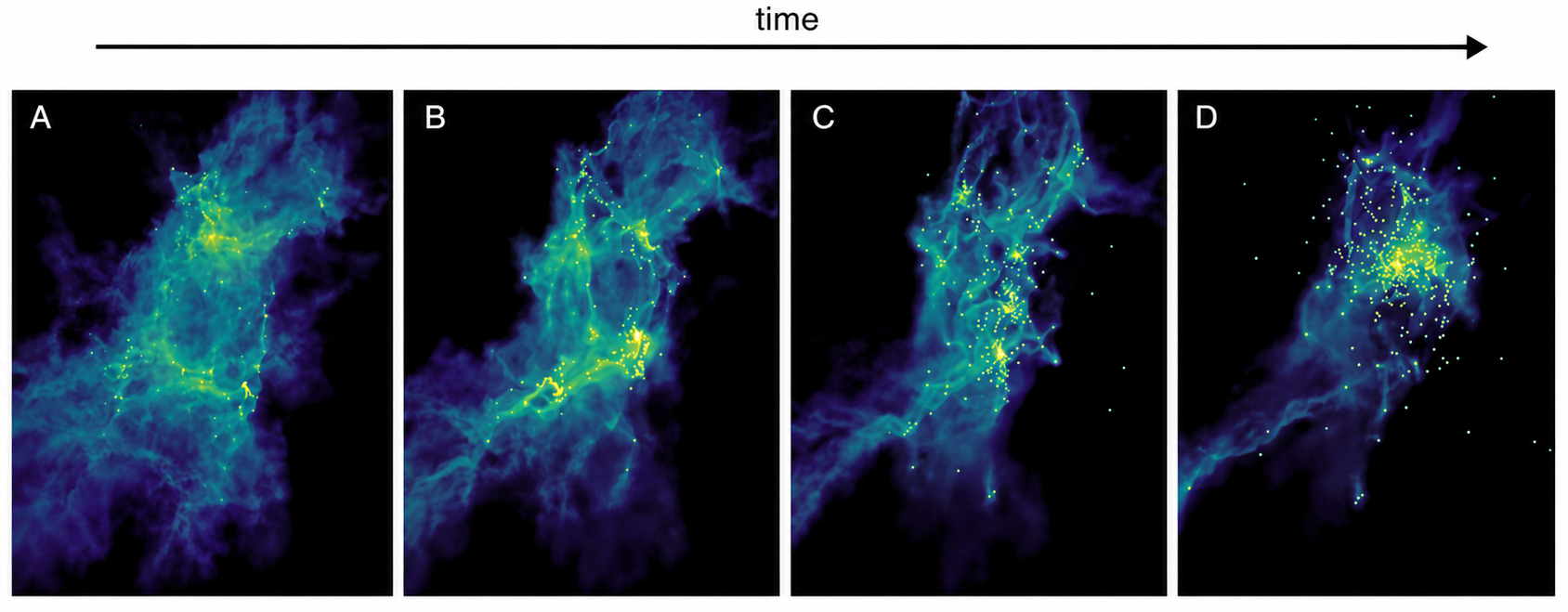}
    \caption{Snapshots of the logarithm of the column density in a 1 pc $\times$ 1 pc region produced from the simulations of \cite{bonnell2003hierarchical-0f2} at four consequent times from left to right ($1.9$, $2.6$, $3.4$, and $4.5$ $\times 10^5$ yr). The color scale spans from a minimum of 0.025 (black) to a maximum of 250 (yellow) g cm$^{-3}$. Each dot indicates a star..}
    \label{fig:Figure_B03}
\end{figure*}

SPH simulations of star formation moved in the direction of a full 3D framework, while still (at least in the early works) maintaining a pure hydrodynamical approach coupled to self-gravity and simple barotropic or isothermal equations of state. The very first 3D calculation to follow the collapse of a molecular cloud core to stellar densities was performed by \cite{bate1998collapse-d05}, followed by several pieces of work focusing on different aspects of the outcome of the star formation process. Within this framework, protostars form in clustered environment (see Figure \ref{fig:Figure_B03}): as opposed to isolated collapse models, these simulations showed that accretion of material is a dynamical, competitive process, often terminated by ejection rather than smoothly declining \citep{bonnell1997accretion-d17, bonnell1998formation-6b4, bonnell2001competitive-20e, bonnell2003hierarchical-0f2, bate2003formation-bc0}; as a consequence, the final mass of a protostar is determined by far more variables than only its initial core properties. 
These calculations demonstrated that non-steady accretion is a generic outcome of the hydrodynamic collapse in clustered environments, pinpointing a physical origin for accretion variability beyond the instability in the disk itself. Subsequent extensions of the SPH simulations included radiative transfer and are discussed in Section\,\ref{subsec:rad_hydro}. 
In parallel to the thin-disk and SPH approaches, grid-based pure hydrodynamic simulations have also been employed in the context of molecular cloud core collapse, although to a smaller extent \citep{matsumoto2003fragmentation-ae3, banerjee2004formation-4f7}. These studies provided an Eulerian validation of the non-steady nature of accretion, in absence of other physical mechanisms, as a natural outcome of gravitational collapse.

The main limitation of simple, purely hydrodynamical models is that they neglect radiation (Section \ref{subsec:rad_hydro}) and magnetic fields (Section \ref{subsec:MHD}). Most early simulations simply modeled the cloud collapse with either an isothermal or barotropic equation of state; the first one holds until densities of $\sim 10^{-13}$ g cm$^{-3}$, after which it would transition to adiabatic \citep{larson1969numerical-ed8, masunaga1998radiation-8e3, masunaga2000radiation-02f}; the adiabatic approximation, however, only works in the assumption that the gas is heating because of compression alone. The barotropic equation of state, on the other hand, is based on the evolution of the temperature at the highest density (as calculated with radiative transfer) during the collapse, and fails to correctly describe the temperature distribution in a complex, three-dimensional simulation. An accurate description of the core fragmentation process (which heavily depends on the gas temperature) however cannot neglect the radiative description of the central protostar: once it forms, the heating of the gas due to accretion vastly exceeds that caused by compression, to the point where it can be sufficient to prevent the fragmentation of low-mass protostellar disks into brown dwarfs \citep{matzner2005protostellar-b9d}. In this context, the natural improvement of simple hydrodynamical models is the coupling with radiative transfer to what is commonly referred to as radiation hydrodynamics (RHD).

\subsection{Radiation Hydrodynamics (RHD)}\label{subsec:rad_hydro}

%The inclusion of radiative transfer in simulations of core collapse allows to account for the radiative feedback of the accreting central protostar, which in turn leads to a far more accurate description of the temperature structure. The fragmentation process and outcome is strongly impacted, as the Jeans mass (and therefore the resulting protostellar mass) depends on the gas temperature; 

\begin{figure}
    \centering
    \includegraphics[width=0.5\linewidth]{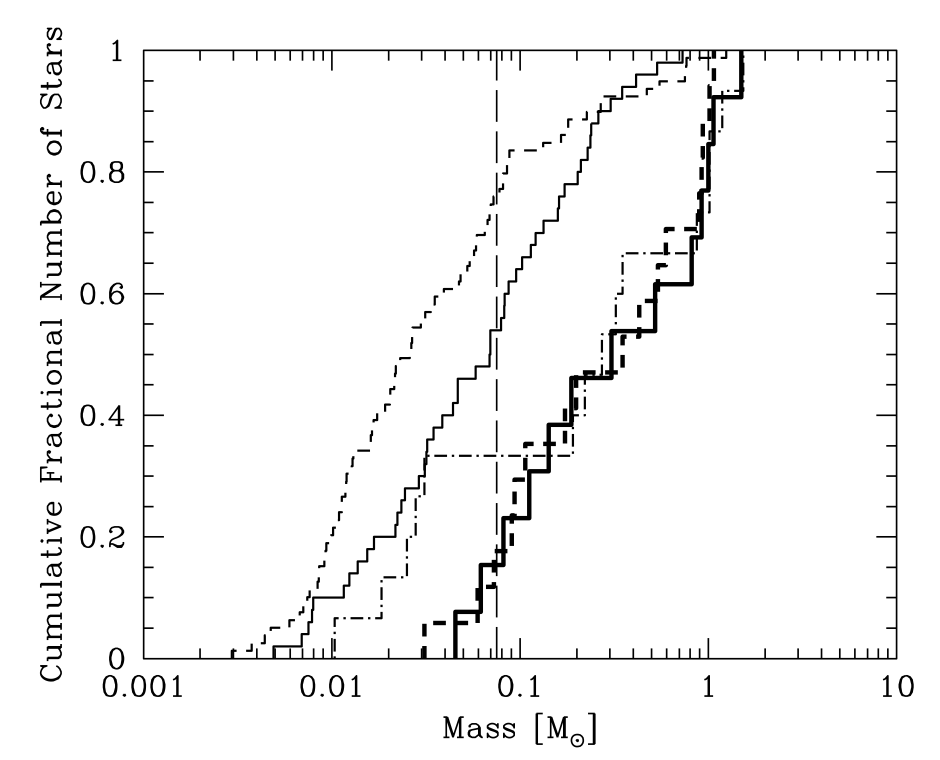}
    \caption{Cumulative initial mass function produced by a barotropic equation of state (thin solid and dashed lines) and radiation hydrodynamical calculations (thick solid, dashed, and thin dot-dashed lines). Adapted from \cite{bate2009c} with permission, CC BY 4.0.}
    \label{fig:B09}
\end{figure}

The addition of radiation transport in simulations of molecular cloud collapse was implemented in both SPH and grid-based codes, often in the flux-limited diffusion (i.e., treating radiation like a diffusing quantity and imposing a physically-motivated limit on the flux) and gray (i.e., integrated over the entire frequency range) approximation. 
One of the first three dimensional implementations by \citep{whitehouse2006thermodynamics-6c1} showed a remarkable dependence in the temperature evolution from the initial conditions, supporting the need for radiative transfer models to obtain an accurate temperature description. Several works \citep{krumholz2007radiation-hydrodynamic-a29, bate2009c, offner2009effects-be6} confirmed the role of radiation feedback from accreting protostars in inhibiting fragmentation, thereby resulting in fewer, more massive objects with higher accretion rates; \cite{bate2009c} estimated the protostar formation efficiency to be reduced by a factor four compared to the hydrodynamical simulations with a barotropic equation of state (see Figure \ref{fig:B09}). Other works have also employed frequency-dependent radiative transfer models (see, e.g., \citealt{vaytet2012simulations-b12, vaytet2013simulations-f88}, which however confirmed the general findings of the gray approximation models, which manage to capture well the key aspects of thermal evolution relevant for the formation of protostars. \cite{bate2010collapse-626} evolved 
RHD simulations beyond the formation of the stellar core, finding that the energy released in the process is high enough to drive a shock wave through the disk, which dramatically decreases the accretion rate on to the stellar core and launches a bipolar outflow.

%The circumstellar disk increases its size and mass until the end of the main accretion phase as shown in Machida et al. (2010a) BUT THIS IS NOT RHD! JUST HYDRO, in which the evolution of the circumstellar disk in the unmagnetized cloud was investigated. 

%However, the radiation pressure does not halt accretion \cite{krumholz2009formation-3e8} this is true but they talk about very high stellar masses like >25 msun

%\cite{bate2009c}

% other early models: Krumholz2006 basically same as W&B06
% other method which computes the mean optical depth: Stamatellos, Whitworth, Bisbas, Goodwin 2007
% Krumholz2007b is AMR, grid-based

%indeed, several works \citep{rafikov2005can-00d, matzner2005protostellar-b9d, kratter2006fragmentation-b5c, whitworth2006minimum-8a5} suggested that an accurate treatment of radiative transfer would likely decrease disk fragmentation.

%\cite{vorobyov2006burst-f6a}: focus on the burst mode of protostellar accretion - prolonged quiescent periods of low accretion rate (typically $\leq 10^{-7}$ M$_{\odot}$ yr$^{-1}$) punctuated by intense bursts of accretion (typically $\geq 10^{-4}$ M$_{\odot}$ yr$^{-1}$, lasting less than 100 yr) during which most of the protostellar mass is accumulated. Conclude that most 

%\cite{vorobyov2005effect-a0d, vorobyov2006burst-f6a, vorobyov2010burst-a5d, vorobyov2015variable-98b, vorobyov2018early-117, vorobyov2020accretion-65a}

\subsection{Magnetohydrodynamics (MHD)}\label{subsec:MHD}

The applicability of the pure hydrodynamical and RHD framework essentially depends on the ionization of the medium. In principle, only a completely unionized medium can be described in a purely hydrodynamical framework; however, stars form from the condensation of galactic interstellar medium (ISM), where the contribution of magnetic fields cannot be neglected. Magnetic fields are ubiquitously observed at all scales of the star-formation process, from molecular clouds \citep{crutcher1999magnetic-0a3, bourke2001new-5ca, soler2017what-0e2} to pre-stellar cores \citep{heiles2005cosmic-c9f, troland2008magnetic-6c7, maury2018magnetically-081} to protoplanetary disks \citep{alves2018magnetic-1e0, ohashi2025observationally-686}, and observational evidence suggests they play a crucial role in shaping key observables including accretion flows, disk masses and sizes \citep{maury2018magnetically-081, galametz2020observational-659, cabedo2023magnetically-bb9}. The interested reader can find dedicated reviews in \cite{crutcher2012magnetic-2ff, hennebelle2019role-da6, krumholz2019role-10a, zhao2020formation-af8, pattle2023magnetic-73e}.

Magnetization is fundamental in the collapse of pre-stellar cores in that it impacts the evolution of angular momentum. While angular momentum is essentially constant in an unmagnetized medium \citep{matsumoto2003fragmentation-ae3}, the presence of magnetic fields and therefore a magnetic tension makes it possible to exchange angular momentum between the fluid particles through Alfvén waves \citep{shu1987star-868, zhao2020formation-af8}; as a consequence, the angular momentum of the core is reduced and the rotation slowed down, a process called magnetic braking, which prevents the formation of unrealistically large disks \citep{hennebelle2016magnetically-535, wurster2018effect-6a2, zhao2020formation-af8, lee2021universal-bdc, lee2024protoplanetary-b9f}.

In the following, we describe the physics of magnetic models qualitatively and refer the reader to the review of \cite{hennebelle2019role-da6} for a detailed mathematical treatment.

\subsubsection{Ideal MHD}

The easiest assumption we can make when introducing magnetic fields is the so-called ideal MHD, which considers fluids as perfect conductors, i.e., perfectly coupled to the field. The main implication of this assumption is that the dissipative terms in the magnetic equations vanish, therefore resulting in a state of constant entropy; the fluid particles are attached to the field lines, so that they can flow along them but cannot cross them, a phenomenon commonly referred to as flux freezing. In the flux freezing regime, the twisting of the magnetic field lines caused by rotation applies a counter force to the rotation itself, effectively slowing rotation down (hence the name magnetic braking) and increasing the infall of gas in the radial direction. 

Magnetic braking plays a fundamental role in reproducing the observed disk sizes; however, in the flux freezing regime, it can be strong enough to fully impede the formation of disks around protostars, the so-called ``magnetic braking catastrophe". 
Indeed, simulations of collapsing cores in the ideal MHD approximation found magnetic braking to efficiently remove angular momentum during the collapse, while simultaneously launching large-scale bipolar outflows, and hinder the formation of rotationally-supported disks \citep{allen2003collapse-b9d, hennebelleteyssier2008, joos2012protostellar-c97, machida2004first-162, machida2005collapse-87e, shu2004does-b5c}. 
3D simulations confirmed the strong suppression of disk formation in the Class\,0 phase, when even allowed at all \citep{price2007impact-219, mellon2008magnetic-621, hennebellefromang2008, galli2006gravitational-193}. 

Even when radiative transfer is included, ideal MHD simulations predict very small or transient disks and efficient angular-momentum extraction (e.g., \citealt{commeron2010protostellar-35e, tomida2010radiation-b74}). 
The ideal approximation holds when the ionization rate $\zeta_i$ is high enough \citep[precisely, $\zeta_i >10^{-13}$ s$^{-1}$,][] {wurster2018effect-6a2}; however, the typical ionization rates of molecular clouds are low \citep[i.e., $\zeta_i < 10^{-16}$ s$^{-1}$,][]{redaelli2021cosmic-ray-3cb, pineda2024probing-a3c} and the commonly assumed value is $\zeta_i = 10^{-17}$ s$^{-1}$ \citep{padovani2009cosmic-ray-a6f, neufeld2017cosmic-ray-25f}. The low ionization rate implies a substantial influence of non-ideal MHD effects, which need to be accounted for in star formation simulations, and are able to prevent the magnetic braking catastrophe \citep{mellon2009magnetic-665, dapp2012bridging-bba, tomida2015radiation-9fe, wurster2016can-1a6, zhao2016protostellar-c4f, vaytet2018protostellar-958, tsukamoto2015effects-682, marchand2018impact-fbd}; however, there are other ways to restore disk formation even in ideal MHD, such as considering misalignment between the rotational axis and the magnetic field \citep{hennebelle2009disk-73e, joos2012protostellar-c97, li2013does-d6e, krumholz2013protostellar-e7f, codella2014alma-c5e, lee2017first-c36} and the influence of turbulence and dynamical environments \citep{machida2011origin-b2e, seifried2012disc-75b, santos-lima2012role-5ee, wurster2019there-661, he2023massive-ecf, he2025formation-3e6, mayer2025protostellar-2fb}.

\subsubsection{Non-ideal MHD}

Moving away from the ideal MHD approximation implies introducing dissipative terms, to describe the fact that not all particles are perfectly coupled to the magnetic field and, therefore, do not have an instantaneous response to changes in the magnetic field itself. The three key effects introduced in non-ideal MHD models are Ohmic dissipation (electron-ion and electron-neutral collisions), ambipolar diffusion (ion-neutral collisions), and the Hall effect (drift velocity between positive and negative ions, dispersive term -- \citealt{wardle2004star-84e}): these account for the finite resistivity and the inertia between the different charge carriers \citep{wardle1999conductivity-a9e, pandey2008hall-59d, zhao2018effect-09f}. The relative importance of the non-ideal effects depends on the coupling of the charged particles with the magnetic field, so that they dominate in different regimes; in the context of protoplanetary disk evolution, \cite{tsukamoto2017impact-515} and \cite{wurster2021do-e2d} have shown that a complete treatment of disk formation ought to include all three.

Ambipolar diffusion, also referred to as the ion-neutral drift, has a particularly important role in star formation. In partially ionized fluids, as is the case for molecular clouds, \cite{mck07, pineda2024probing-a3c}, the magnetic field can have an indirect impact on neutral particles through their collisions with the ions and the consequent exchange of momentum. In general, compared to the unmagnetized case, the collapse of a magnetically supported core is significantly slowed down; as the neutrals cross the field lines, the magnetic flux decreases and eventually becomes low enough that the gravitational collapse can take place. Several simulations \citep{basu2004formation-f3d, heitsch2004turbulent-728, li2004magnetically-ef4, nakamura2008magnetically-3fc, nakamura2011clustered-39d, loo2008effect-f35, vzquez-semadeni2011molecular-106, bailey2014non-ideal-1e0, bailey2017ionisation-0a7} have shown that subcritical magnetic fields (i.e., giving a high enough magnetic support to inhibit the dynamical collapse) drastically reduce the star formation rate. 

\subsubsection{Large-scale simulations}\label{subsubsec:largescale}

\begin{figure*}
    \centering
    \includegraphics[width=0.97\linewidth]{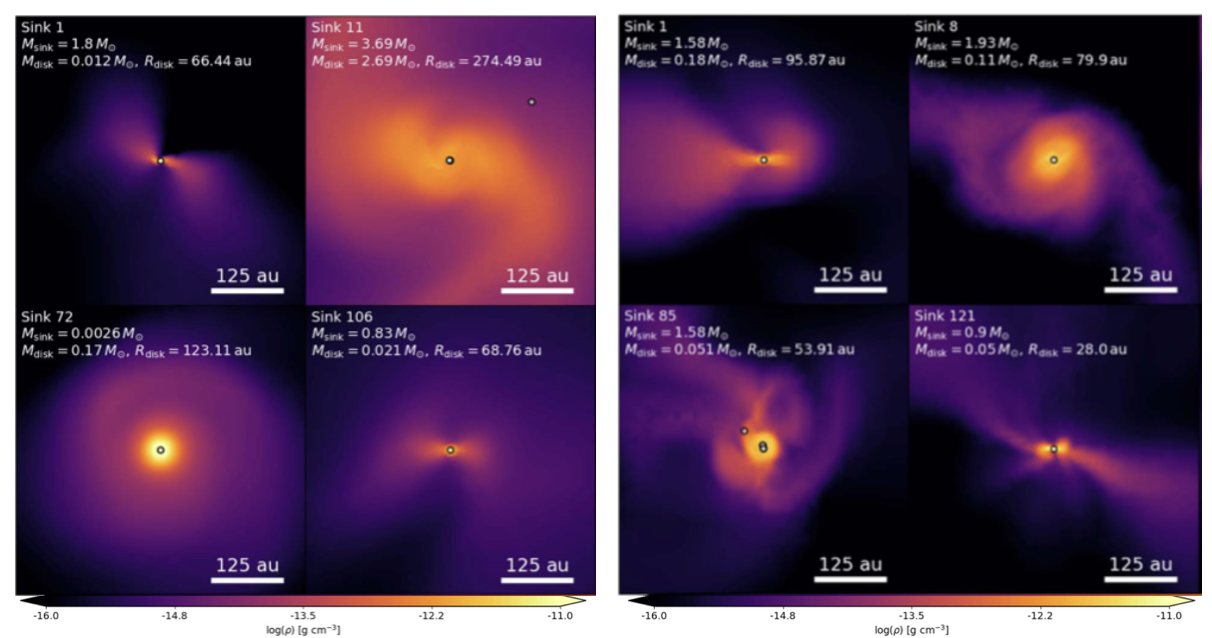}
    \caption{Density maps centered around selected sink particles (indicated by the black circles) in ideal (left) and non-ideal (right) MHD core-collapse simulations after 117 kyr of evolution. Adapted from \cite{Lebreuilly2021}. © AAS. Reproduced with permission.}
    \label{fig:Ugos_sims}
\end{figure*}

A realistic treatment of the large- (i.e., clump-) scale magnetic field is also fundamental to produce consistent protoplanetary disk populations. Suites of simulations in both the RHD \citep{elsender2021statistical-dbf}, ideal \citep{kuffmeier2017zoom-in-3ed, kuffmeier2019bridge-42d, bate2018diversity-0a3} and non-ideal \cite{wurster2020non-ideal-87f, Lebreuilly2021, Lebreuilly2024, lebreuilly2024influence-984} MHD framework have been run to gain a statistical insight, with a zoom-in technique allowing to resolve the smaller disk scales. Figure \ref{fig:Ugos_sims} shows an example of the output of these models \citep[adapted from][]{Lebreuilly2021}: discs form around sink particles within the simulation, with different morphology and characteristics depending on the physics considered. Zooming-in on each sink, it is possible to determine key disc properties such as the mass and radius; moreover, signatures of the accretion process are visible in the flow of material around and within the disc.

Recently, \cite{ahmad2025birth-8b6} have modeled the formation of protostars and disks finding that magnetic fields and turbulence drive highly anisotropic accretion onto the disk via dense streamers. 
MHD instabilities have also been found to shape the infalling envelope into streamers rather than axially symmetric infall \citep{machida2025complex-bec}; note that the origin of streamers is not limited to instabilities, but can be the consequence of pre-existing turbulence \citep{tu2024fragmentation-e15} or misalignment between the magnetic and rotation axes \citep{machida2026twisted-e3e}.

\section{Protostellar accretion: the emerging picture}
\label{sect:main-res-open-questions}

\subsection{Uncertainties in the retrieval and comparison of observables from simulations}\label{subsec:obs_from_sims}

The availability of a large number of numerical simulations enables statistical analyses and the identification of robust trends. A comparison between numerical results and observations is both natural and ultimately desirable; however, such comparisons must be approached with care to ensure that physically equivalent quantities are being contrasted. Establishing a fair correspondence between simulated and observed properties is non-trivial for both physical and numerical reasons, which we discuss below.

\subsubsection{Stellar properties and mass accretion rate} 

The main limitation to accessing the central star in simulations is the numerical resolution. This is especially easy to grasp for simulations employing sink particles, where the undeniable advantage of maintaining a reasonable computational time is exchanged for the introduction of unresolved regions around the protostars\footnote{Note that the sub-resolution issue is in principle independent on the use of sink particles; we just used them as an example for ease of understanding.}. Furthermore, as opposed to the observational determination of accretion rates -- which is essentially a luminosity measure, as described in Section\,\ref{sect:observations}, numerical simulations often calculate the actual amount of mass transferring from the disk/envelope system onto the innermost region within the resolution limit. 
The caveat is therefore twofold: for one, the actual diagnostic is different -- the result of a calculation based on the stellar parameters and some luminosity on one hand, and an actual measure of accreting mass on the other; furthermore, the accretion rate as measured in simulation is that \textit{onto the smallest resolution element} in the proximity of the star, not necessarily the star itself. If this is the case, then the accretion rate would be a combination of the material reaching the actual protostar and the surrounding region, within the resolution limit, somewhat corresponding to the inner disk.

%RAMSES \cite{teyssier2002cosmological-a00}: also \cite{kuffmeier2018episodic-0c6} use it and show that the environment determines the accretion profile of the protostar. After that, they use MESA to simulate the impact of the accretion outbursts on the spectral energy distribution, showing that there are observable enhancements between 20 and 200 $\mu$m in Class 0 objects.

\subsubsection{Accretion luminosity}

The observable of accretion is the accretion luminosity $L_{\rm{acc}}$ (Section\,\ref{sect:observations}), and the mass accretion rate is then computed by combining $L_{\rm{acc}}$ with the stellar properties and inverting Eq.\,\eqref{eq:01}. However, protostellar $L_{\rm{acc}}$ is not directly observed, but rather derived %from the $L_{\rm{acc}}-L_{\rm{line}}$ relation 
(see Section\,\ref{sect:empirical-relations}).

On the contrary in most numerical simulations, such as those by Bate or Lebreuilly, the path is reversed. The accretion rate is determined from the (radiation-, magneto-)hydrodynamics of the gas, as the flow of material reaching the protostar (or sink particle), and $\lacc$ is then computed based on Eq.\,\eqref{eq:01}. Even accounting for radiative feedback, the caveat persists: if the $\lacc$ is derived from $\macc$, adding a consistent treatment of radiation will ``only" ensure a correct impact on the rest of the structure, but will not remove the fundamental assumption that luminosity is a derived quantity. As these large-scales simulations often do not resolve the transport of angular momentum from the inner disk to the stellar surface, the $\dot M_{\rm{acc}}$ - $\lacc$ conversion requires assuming (i) a protostellar radius and (ii) the radiative efficiency (see Section \ref{sect:observations}); the physics of gas transport in the inner disc and from the disc to the star is also often not represented in large-scale simulations. Finally, determining the confidence level of the derived accretion luminosity is not trivial, and is often based on statistics (for example, it could be defined as the standard deviation in the measurements obtained for a population of objects) or outburst amplitudes within the model rather than an uncertainty in the value per se.

From an observational perspective, the accretion luminosity represents the primary measurable quantity, whereas the mass accretion rate is a derived parameter that inherits and amplifies the uncertainties associated with $L_{\rm acc}$ and with the stellar properties entering Eq.\,\eqref{eq:01}. As a consequence, observational estimates of $\dot M_{\rm acc}$ typically carry substantially larger uncertainties than those on $L_{\rm acc}$. This introduces an additional caveat when comparing simulations and observations: an apparent agreement in $\dot M_{\rm acc}$ may partly reflect the large propagated uncertainties affecting the observationally inferred accretion rates, rather than a genuine consistency between models and data. 

There are also numerical studies, like in \citealt{myers1998evolution-a66}, which instead \textit{prescribe} an accretion history based on analytical models (note that these are, indeed, closer to models than full large-scale simulations); in this case, the accretion luminosity is not derived from the accretion rate anymore, but is a direct result of the assumed model rather than the output of a self-consistent simulation. Observed and simulated accretion luminosities and mass accretion rates should therefore be compared with particular care, taking into account the different assumptions and sources of uncertainty underlying their derivation.

\subsubsection{Synthetic observations}\label{subsubsec:synthetic_obs}

A promising approach to exploit the predictions from simulations is producing synthetic observations. Synthetic observations (see \citealt{haworth2018synthetic-5b7} for a review) are essentially radiative-transfer post-processed results of numerical simulations, whose outputs can be directly compared with the observational data (such as continuum images or SEDs). The advantage of this method is that going from a physical model to synthetic observables usually require fewer assumptions than converting an observation into a physical model; on the other hand, it is important that the post-processing includes the instrumental effects and observational biases specific to the dataset to compare against (e.g., \citealt{koepferl2017fluxcompensator-8e3}), to ensure one-to-one correspondence. The production of synthetic observations, and the extraction of physical observables with the same techniques employed with real data, ensures a fair comparison beyond the naive assumption that simulations and observations can access the exact same quantity (although, a fair comparison still does not remove the uncertainties introduced below the synthetic observation overlay). In the last couple of decades, several studies have performed post-processing of simulations across the whole star formation process, from the large scales of the ISM (e.g., \citealt{smith2014co-dark-e4d, juvela2019synthetic-405}), to the intermediate clump fragmentation \citep{nucara2025rosetta-019}, to the smallest cores and disks (e.g., \citealt{dipierro2015planet-0d7, maury2022recent-4a3, redaelli2024testing-82a}). 

Another perk of synthetic observations is that they allow to test the accuracy of the methods employed to derive physical properties from observational data. In the context of Class 0/I objects, \cite{tung2024accuracy-190} performed synthetic observations of the population of YSOs produced by the non-ideal MHD simulations of \cite{Lebreuilly2024}, with the goal of comparing the retrieved disk radii and masses to the real\footnote{Note that, in this context, by 'real' we mean simulated as the starting point is a numerical simulation.} values. They found that, while disk sizes can be accurately recovered (with uncertainty of a factor 1.6--2.2), the same does not hold for the disk masses. Indeed, disk masses are systematically underestimated (from a factor $\sim 2$ to 10 in the most extreme cases), because of the failure of the assumption of optically thin emission - fundamental in the conversion of flux into mass.

Accretion-related quantities pose an even greater challenge. To date, synthetic observations aimed at recovering accretion luminosities or mass accretion rates remain very limited, as both $L_{\rm acc}$ and $\dot M_{\rm acc}$ depend on sub-grid accretion physics and observational diagnostics that are not self-consistently captured in current post-processing frameworks, and are therefore typically prescribed rather than inferred in a forward-modeling sense.

\subsection{Discrepancies in accretion luminosity estimates from observations}
\label{sect:lacc-discrepancy}

A long-standing issue in the study of protostellar accretion concerns the wide range of accretion luminosities inferred for Class\,I sources by different observational studies \citep[see also,][]{Fischer2023}.  
While it is generally expected that protostars accrete at higher rates than CTTSs, early works based on NIR diagnostics suggested comparable accretion levels for Class\,I and Class\,II objects. In particular, \citet{muz98} and later \citet{Beck2007} found accretion luminosities for Class\,I protostars similar to those measured in CTTSs, a result that was interpreted as evidence for episodic accretion or for most of the stellar mass being assembled at earlier evolutionary stages.
However, subsequent studies based on larger samples and different diagnostics have challenged this picture, reporting systematically higher $\lacc$ and $\macc$ in Class\,I sources compared to Class\,II stars \citep[e.g.,][]{nis05b, Fiorellino2021, Fiorellino2023}. 

One potential source of discrepancy lies in the treatment of extinction. For instance, \citet{Beck2007} derived extinction values assuming that intrinsic hydrogen line ratios follow Case\,B recombination predictions \citep{Hummer1987}, i.e., optically thick conditions in the Lyman lines and optically thin emission in higher-order transitions.
If this assumption does not hold in dense, optically thick protostellar environments, the resulting extinction corrections, and hence the inferred accretion luminosities, may be underestimated.
In addition to methodological differences, sample selection effects are likely to play a major role. The Class\,I population is intrinsically heterogeneous, spanning a wide range of envelope masses, accretion histories, and evolutionary stages. Both observations and theoretical models indicate that several physical properties of Class\,I sources, including $M_{\rm env}$, $M_{\rm disk}$, and $\macc$, exhibit dispersions approaching two orders of magnitude \citep[e.g.,][]{Jorgensen2009, Dunham2014, Tobin2015, Fiorellino2023}, naturally leading to a broad distribution of accretion luminosities. Differences in the fraction of sources for which the accretion luminosity dominates the bolometric output further complicate direct comparisons between studies. 
For example, the high fraction (95\%) of sources with $\lacc/\lbol > 0.5$ reported %in the young and actively accreting NGC~1333 cluster \citep{Fiorellino2021} and 
in the mid-infrared–bright sample analyzed by \citet{Testi2025} may reflect the younger age and more embedded nature of these populations. In contrast, the samples analyzed by \citet{Fiorellino2021, Fiorellino2023} and \citet{whi04} show significantly lower fractions of such objects, ranging from 25\% to 40\%. A plausible explanation for this difference is related to sample selection: \citet{Fiorellino2023} considered a deliberately diverse sample of Class\,I sources, while the \citet{whi04} sample was optically visible and therefore likely biased toward less embedded and, on average, lower-accreting objects, potentially including sources at more advanced evolutionary stages.
However, differences in the methodologies adopted to estimate $A_V$ and $\lacc$, particularly between NIR and MIR approaches where the extinction law varies, must also be taken into account when interpreting these results. 
Recent results further highlight the potential impact of diagnostic choices. 
For instance, \citet{Fiorellino2025} showed that, for CTTSs, accretion luminosities derived from Br$\gamma$ systematically trace the lower end of the $\lacc$ distribution, with Br$\gamma$-based estimates representing a lower limit in about 80\% of the cases. 
If a similar effect applies to Class\,I sources or more deeply embedded protostars, it would imply that current estimates of $\lacc$ based on near-infrared tracers may sistematically underestimate the accretion luminosity. This effect could contribute to part of the discrepancy between studies based on different tracers.

Both observational biases and analysis techniques can significantly affect inferred accretion properties. Ultimately, only large, homogeneous samples analyzed using multiple, independent diagnostics and consistent methodologies will allow us to disentangle intrinsic source-to-source variations from systematic effects, and to establish robust constraints on protostellar accretion luminosities. Applicability and limitations of empirical relationships are discussed in Section\,\ref{sect:empirical-rel-discussion}

\subsection{Protostellar mass budget}

\begin{figure}
    \centering
    \includegraphics[width=0.86\linewidth]{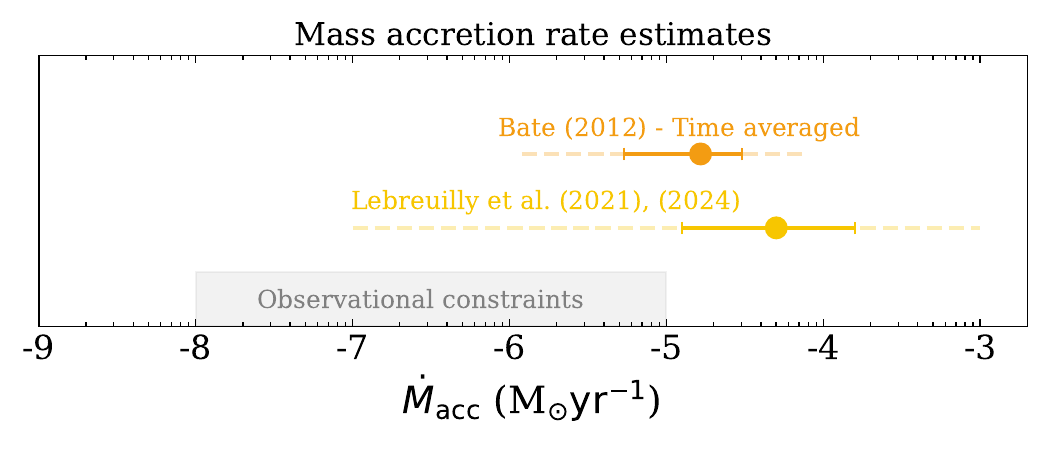}
    \caption{Comparison of the observed range of values (gray shaded region) with simulation-inferred mass accretion rates from two numerical setups. We indicate the mean value with a circle and the 16th and 84th percentile ranges with a solid line; the dashed lines represent the interval between the minimum and maximum values.}
    \label{fig:Macc-comparison}
\end{figure}

Notwithstanding the caveats discussed in Section\,\ref{sect:lacc-discrepancy}, comparing the simulated $L_{\rm acc}$ and $\dot M_{\rm acc}$ with the observational constraints is still informative. 
While there are no accretion surveys of Class\,0 YSOs, the typical accretion luminosities and mass accretion rates ranges observed in Class\,I and FS sources are about $-2.5 < \log(\lacc/\lsun) < 0.5$ and $-8 < \log(\dot M_{\rm{acc}}/$M$_{\odot}\rm{yr}^{-1}) < -5$ \citep{fiorellino2023mass-9cb}. 

From the numerical perspective, to obtain a fair comparison with observational surveys, it is necessary to consider works focused on producing statistically significant samples. Most of the currently available literature on simulations of protostellar accretion, however, does not satisfy this requirement: the majority of works simulate single sources, or a handful (like the case of \citealt{kuffmeier2018episodic-0c6}, with six sinks in total), but not enough to derive meaningful statistical properties. Large-scale simulations certainly have a higher computational cost, but remain indispensable to correctly interpret the observational results: at the time of writing, the only published works that can be considered for this scope are those of \cite{bate2012stellar-086} and \cite{Lebreuilly2021, lebreuilly2024influence-984}.

Qualitatively speaking, these models recover the same orders of magnitude and reach higher rates; Figure\,\ref{fig:Macc-comparison} shows the ranges of derived accretion rates across different simulations, spanning from around $10^{-9}$ to $10^{-4}$ M$_{\odot}\rm{yr}^{-1}$, compared to the observational constraints. 
A possible interpretation of the higher simulated accretion rates is that they can only be sustained during the Class\,0 stage, possibly accreting via non-magnetospheric accretion scenarios. 
To test this possibility, developing new robust methodologies to constrain accretion rates on Class\,0 is fundamental. Similarly, it is currently very challenging to infer $\macc$ during FUors-like bursts, because the hot disk covers the information on the stellar photosphere, making very challenging to determine the stellar radius and mass, needed to compute $\macc$.

If accretion rates of $\dot{M}_{\rm acc} \sim 10^{-5}$--$10^{-4}\,M_\odot\,{\rm yr^{-1}}$ are confirmed in very embedded sources (e.g., Class\,0 and the faintest Class\,I objects), and if sustained over the duration of the protostellar phase (of order $\sim$1~Myr), they would be sufficient to assemble low-mass stars. 
However, evolutionary models in which the inner envelope mass declines exponentially predict infall rates in the range $\sim10^{-6}$--$2.6\times10^{-5}\,M_\odot{\rm yr^{-1}}$ during the Class\,0 phase to form stars between 0.1 and 2\,M$_\odot$ \citep[e.g.,][]{Fischer2017}. These values suggest that extremely high sustained accretion rates (e.g., $\sim10^{-4}\ M_\odot\,{\rm yr^{-1}}$), which are not supported by typical observed bolometric luminosities, may not be required.
Conversely, if typical accretion rates do not exceed $\sim 10^{-6}\,M_\odot\,{\rm yr^{-1}}$, an additional contribution to the mass budget would likely be required, possibly in the form of episodic high-accretion events such as FU~Orionis–like outbursts. 
\citet{Zakri2022} estimate of burst frequencies in Class\,0 protostars of order one event every $\sim$400\,yr indicate that a non-negligible fraction of the final stellar mass could, in principle, be accreted during such episodes.
Therefore, the relative contribution of steady versus episodic accretion to the global mass assembly remains an open question, strongly dependent on the true burst duty cycle and on the time-averaged accretion history.

A further caveat in protostellar accretion studies is that many key physical parameters are not directly measured, but inferred from observational tracers through model-dependent assumptions. Accretion luminosities are typically derived from emission-line fluxes via method based on empirical relations, with intrinsic dispersions of order $\sim$0.3\,dex, and are additionally affected by uncertainties in extinction (often $\sim$1\,mag), veiling, and the physical origin of the line emission. 
Stellar masses and radii are commonly estimated by placing sources on PMS evolutionary tracks, which remain uncertain during the protostellar phase owing to ongoing accretion and poorly constrained initial conditions, potentially introducing uncertainties up to $\sim$1\,dex. These combined effects propagate into the derived mass accretion rates, for which total uncertainties can approach $\sim$1.5$-$2\,dex.

\subsection{How do protostars accrete?}
\label{sect:accretion-nature}

Given the challenges in observing and modeling the protostellar phase, the nature of the protostellar accretion process remains an open question. The first natural hypothesis is that protostellar accretion follows the magnetospheric accretion paradigm, well established for CTTSs. However, several key aspects must be considered.

\begin{itemize}
    \item The strength and topology of stellar magnetic fields in protostars are poorly constrained. While kilo-Gauss magnetic fields are routinely measured in CTTSs \citep[e.g.,][]{JohnsKrull2007, Donati2011, PerezPaolino2025}, the high extinction affecting embedded protostars prevents direct measurements of their photospheric emission, making it extremely challenging to determine their magnetic field properties. A sample of Class\,I and FS protostars shows the same $B$ distribution as CTTSs \citep{flores2024}, while the sample from \citet{Drouglazet2026} reports also some non detected magnetic field, pointing to a diversity in magnetic field topology and accretion geometry within the Class\,I and FS population, often (although not always) strong enough to sustain the magnetospheric accretion scenario.

    \item Low-mass protostars are expected to be larger than more evolved PMS stars, with typical radii $R_\star \sim 3$--$5\,R_\odot$ according to protostellar evolutionary models \citep{Larson1969, StahlerPalla2004}, and are likely fully convective. As a consequence, their dynamo mechanisms is expected to operate in a different regime, potentially leading to magnetic fields with distinct topology and time variability compared to those of more evolved CTTSs \citep[e.g.,][]{Chabrier2006, Browning2008, Gregory2012}, which may result in magnetic field configurations that differ in strength and topology from those observed in more evolved CTTSs. %may be less efficient or produce more complex and time-variable magnetic field configurations than those observed in CTTSs. 

\item Protostellar disks are more massive, geometrically thicker, hotter, and likely more turbulent than CTTS disks due to continuous mass loading from the envelope. The evidence for systematically higher disk masses, however, is mixed and may depend on dust modeling assumptions \citep[e.g.,][]{Sheehan2022, Tobin2024}. Such conditions may hinder the formation of a stable magnetospheric cavity and the development of ordered funnel flows.

\item Protostellar mass accretion rates must be significantly higher than those observed in CTTSs in order to build the final stellar mass within the first $\sim$1~Myr after core collapse. 
Typical mass accretion rates measured in CTTSs, $\macc \sim 10^{-11}$--$10^{-7}\,M_\odot\,\mathrm{yr^{-1}}$, integrated over a Class~II lifetime of $\sim$2~Myr \citep{Evans2009}, imply that only a minor fraction of the final stellar mass can be assembled during the Class~II phase. Most of the stellar mass must therefore be accreted earlier, during the embedded stages, requiring typical protostellar accretion rates of $\macc \sim 10^{-6}$--$10^{-4}\,M_\odot\,\mathrm{yr^{-1}}$.

Observationally inferred protostellar accretion rates are often lower than these values (see Section~\ref{sect:empirical-relations}), but this apparent discrepancy should be interpreted with caution. Current estimates ({\it i}) are generally derived under the assumption of magnetospheric accretion, whose applicability during the earliest protostellar stages remains uncertain or are determined by assuming $\lacc \sim \lbol$, providing only upper limits; ({\it ii}) are available primarily for Class\,I and FS objects, which trace relatively evolved phases and may accrete at lower rates than Class\,0 sources \citep[e.g.,][]{Fischer2017, Laos2021, LeGouellec2024}; and ({\it iii}) are biased toward the brightest and least embedded objects, preferentially selecting more evolved systems with potentially lower accretion activity. As a consequence, currently measured protostellar accretion rates may not capture the full mass assembly history. An alternative, and not mutually exclusive, explanation is that a substantial fraction of the stellar mass is accreted through episodic bursts, which are largely missed by time-averaged diagnostics \citep[e.g.,][]{Audard2014, Vorobyov2015, Zakri2022}.

From a theoretical standpoint, high accretion rates challenge the standard magnetospheric picture. The truncation radius scales as $R_{\rm m} \propto B^{4/7}\dot{M}_{\rm acc}^{-2/7}$ \citep{Koenigl1991}, implying that even magnetic field strengths comparable to those measured in CTTSs would yield a substantially reduced $R_{\rm m}$ at protostellar $\macc$, potentially approaching $R_\star$. In this regime, magnetospheric funnel flows and high-latitude accretion shocks may become inefficient or observationally indistinguishable from direct disk--star accretion. 
One plausible alternative is boundary-layer accretion, in which the disk material accretes directly onto the stellar surface without being channeled by a large-scale stellar magnetosphere. Rather than being truncated at several stellar radii, the inner disk extends down to the stellar surface, where the gas dissipates its excess angular momentum and kinetic energy within a narrow boundary layer at the star–disk interface \citep{LyndenBell1974, Pringle1981, Popham1993, Popham1997}. 
Such a regime is expected to occur when the stellar magnetic field is too weak, too complex, or overwhelmed by high mass accretion rates to efficiently truncate the disk, conditions that may be common during the earliest protostellar phases \citep{Hartmann2016}. 

\end{itemize}

Taken together, these considerations indicate that the accretion geometry in protostars may differ substantially from that of CTTSs, and that multiple accretion regimes may coexist or operate at different evolutionary stages. 
One possible scenario is that boundary-layer accretion dominates during the earliest, deeply embedded phases, and gradually transitions to magnetospheric accretion at later stages, with the two modes potentially coexisting between the Class\,0 and Class\,I stages.

\section{Conclusions and next steps}\label{sect:conclusions}

Protostellar accretion remains one of the least constrained phases of low-mass star formation, despite its central role in setting stellar masses, disk properties, and the initial conditions for planet formation. Throughout this review, we have shown that this difficulty primarily stems from the intrinsic complexity and time variability of the accretion process, combined with the indirect and often time-averaged nature of the available diagnostics, and with the deeply embedded nature of protostars during their main accretion phase.

A first key result emerging from current observations is that accretion during the embedded phases cannot be described as a steady process. Variability is ubiquitous, and episodic accretion is strongly suggested by both numerical models and time-domain observations. However, while accretion variability is now well established qualitatively, its quantitative role in the protostellar mass budget remains poorly constrained. 
The frequency, duty cycle, intensity, and contribution of accretion bursts to the total assembled mass are still uncertain, particularly for Class\,0 sources, which remain largely inaccessible to direct accretion diagnostics. Ambitious spectroscopic variability survey in the infrared are necessary to quantify the contribution of protostellar variability to the stellar mass budget.

A second major limitation lies in our ability to infer accretion properties from observations. Accretion luminosities and mass accretion rates in protostars are not measured directly, but inferred through empirical relations, assumptions on the accretion geometry, and poorly constrained stellar parameters. Many of these diagnostics were developed for Class\,II systems and implicitly assume magnetospheric accretion, whose applicability during the protostellar phase remains uncertain. As a result, current estimates of $\lacc$ and $\dot M_{\rm acc}$ for embedded sources are subject to large and often difficult-to-quantify systematic uncertainties.

Similarly, disk and envelope properties—key ingredients in regulating accretion—are challenging to measure reliably during the embedded phases. While disk radii can now be constrained with increasing confidence, disk mass estimates remain highly uncertain due to optical depth effects, envelope contamination, and poorly constrained dust properties. These uncertainties directly impact attempts to link accretion rates, disk evolution, and instability-driven variability. Post-processing data analysis specifically applied to the earliest stages could enlight the best measurement strategy.

From a theoretical perspective, numerical simulations robustly predict time-variable and often episodic accretion histories, but comparing these predictions with observations remains non-trivial. Simulated accretion rates, luminosities, and evolutionary timescales do not map straightforwardly onto observationally inferred quantities, and differences in definitions, resolution, and diagnostics complicate direct comparisons. Without a consistent framework to connect physical accretion histories to observable quantities, it remains difficult to assess whether models and observations are genuinely consistent or merely overlapping within large uncertainties. Furthermore, large-scale numerical frameworks are rather limited: with only a handful of studies producing statistically significant samples of synthetic protostars, the derivation of meaningful expected trends and subsequent comparison to observational evidence remains limited.

In this context, progress in understanding protostellar accretion requires a shift in focus. Rather than relying on individual objects or time-averaged diagnostics, future efforts must adopt a population-based approach that explicitly accounts for accretion variability. On the theoretical side, population synthesis models that include episodic accretion, disk evolution, and envelope replenishment are needed to link physical accretion processes to statistically meaningful, observable quantities: such models provide a natural framework to explore how different accretion regimes contribute to stellar mass assembly and to predict the distributions of accretion-related observables. 
Despite the significant computational cost, this development is as a timely, necessary effort to move towards a comprehensive understanding of protostellar accretion. In complementary manner, on the observational side, constraining protostellar accretion demands larger and more homogeneous samples of Class\,0 and Class\,I sources, combined with systematic time-domain studies capable of quantifying accretion variability and burst frequencies. Infrared and submillimeter monitoring, together with coordinated spectroscopic follow-up, will be essential to bridge the gap between instantaneous diagnostics and long-term accretion histories.

Ultimately, protostellar accretion should be regarded not as a single, well-defined mechanism, but as a time-dependent process shaped by evolving disk–envelope structures, variable mass supply, and potentially multiple accretion regimes. Clarifying how these ingredients combine to assemble stellar mass remains a fundamental challenge. Addressing it will require integrating observations and models within a consistent, population-based framework that treats variability as an essential feature of protostellar accretion rather than as a secondary complication. Because this same process sets the mass reservoir, thermal structure, and dynamical conditions of young disks and may operate through analogous mechanisms during planet growth, understanding protostellar accretion is central not only to stellar assembly, but to the physical origin of planetary systems.

The advent of high-sensitivity observational facilities such as the Square Kilometer Array Observatory (SKAO) and the Extremely Large Telescope (ELT) will enable the detailed and accurate study of the most embedded and further ($\lesssim 2$\,kpc) protostars, providing more complete samples and, overall, better statistical constraints. 
In particular, the ELT will allow us to extend our understanding of the star formation process to environments with metallicities different from that of the solar neighborhood. In this context, extending Gaia-like astrometric capabilities to the infrared, as envisioned for GaiaNIR, will be crucial to probe deeply embedded populations and fully characterize the earliest stages of star formation.

\section*{Conflict of Interest Statement}
%All financial, commercial or other relationships that might be perceived by the academic community as representing a potential conflict of interest must be disclosed. If no such relationship exists, authors will be asked to confirm the following statement: 

The authors declare that the research was conducted in the absence of any commercial or financial relationships that could be construed as a potential conflict of interest.

\section*{Author Contributions}

EF and AS were both responsible for writing and editing the manuscript.

\section*{Funding}
%Details of all funding sources should be provided, including grant numbers if applicable. Please ensure to add all necessary funding information, as after publication this is no longer possible.
This project has received funding from the European Research Council (ERC) via the ERC Synergy Grant ECOGAL (grant 855130). Views and opinions expressed are however those of the author(s) only and do not necessarily reflect those of the European Union or the European Research Council Executive Agency. Neither the European Union nor the granting authority can be held responsible for them.

\section*{Acknowledgments}
We acknowledge helpful comments from Ignacio Mendigut{\'\i}a, Doug Johnstone, Tom Megeath, and Lynne Hillenbrand. Their feedback helped us to improve this manuscript.

%\section*{Supplemental Data}
% \href{http://home.frontiersin.org/about/author-guidelines#SupplementaryMaterial}{Supplementary Material} should be uploaded separately on submission, if there are Supplementary Figures, please include the caption in the same file as the figure. LaTeX Supplementary Material templates can be found in the Frontiers LaTeX folder.

%\section*{Data Availability Statement}
%The datasets [GENERATED/ANALYZED] for this study can be found in the [NAME OF REPOSITORY] [LINK].
% Please see the availability of data guidelines for more information, at https://www.frontiersin.org/about/author-guidelines#AvailabilityofData

\bibliographystyle{Frontiers-Harvard} %  Many Frontiers journals use the Harvard referencing system (Author-date), to find the style and resources for the journal you are submitting to: https://zendesk.frontiersin.org/hc/en-us/articles/360017860337-Frontiers-Reference-Styles-by-Journal. For Humanities and Social Sciences articles please include page numbers in the in-text citations 
%\bibliographystyle{Frontiers-Vancouver} % Many Frontiers journals use the numbered referencing system, to find the style and resources for the journal you are submitting to: https://zendesk.frontiersin.org/hc/en-us/articles/360017860337-Frontiers-Reference-Styles-by-Journal
%\bibliography{bibliografia, biblio}
\bibliography{Bibliografia_Combined}

%%% Make sure to upload the bib file along with the tex file and PDF
%%% Please see the test.bib file for some examples of references

%\section*{Figure captions}

%%% Please be aware that for original research articles we only permit a combined number of 15 figures and tables, one figure with multiple subfigures will count as only one figure.
%%% Use this if adding the figures directly in the mansucript, if so, please remember to also upload the files when submitting your article
%%% There is no need for adding the file termination, as long as you indicate where the file is saved. In the examples below the files (logo1.eps and logos.eps) are in the Frontiers LaTeX folder
%%% If using *.tif files convert them to .jpg or .png
%%%  NB logo1.eps is required in the path in order to correctly compile front page header %%%

%%% If you don't add the figures in the LaTeX files, please upload them when submitting the article.
%%% Frontiers will add the figures at the end of the provisional pdf automatically
%%% The use of LaTeX coding to draw Diagrams/Figures/Structures should be avoided. They should be external callouts including graphics.

\end{document}